\documentclass{iopart}

\usepackage{iopams}

\def\a{\alpha}
\def\b{\beta}

\def\phi{\varphi}
\def\la{\lambda}

\def\s{\sigma}
\def\z{\zeta}

\def\C{{\mathcal C}}

\def\D{{\mathcal D}}
\def\E{{\mathcal E}}

\def\G{{\mathcal G}}

\def\R{{\bf R}}

\def\S{{\mathcal S}}
\def\T{{\rm T}}

\def\X{{\mathcal X}}
\def\Y{{\mathcal Y}}

\def\La{\Lambda}

\def\pa{\partial}

\def\d{{\rm d}}       

\def\x{\times}

\def\o+{\oplus}

\def\ss{\subset}
\def\sse{\subseteq}

\def\<{\langle}
\def\>{\rangle}

\def\EOR{\hfill {$\odot$}}
\def\EOP{\hfill {$\triangle$}}

\def\interno{\hskip 2pt \vbox{\hbox{\vbox to .18
truecm{\vfill\hbox to .25 truecm
{\hfill\hfill}\vfill}\vrule}\hrule}\hskip 2 pt}

\def\({\left(}
\def\){\right)}
\def\[{\left[}
\def\]{\right]}
\def\=#1{\overline #1}

\def\~#1{\widetilde #1}
\def\wt#1{\widetilde #1}
\def\.#1{\dot #1}
\def\^#1{\widehat #1}

\def\mapright#1{\smash{\mathop{\longrightarrow}\limits^{#1}}}
\def\mapdown#1{\Big\downarrow\rlap{$\vcenter{\hbox{$\scriptstyle#1$}}$}}
\def\mapleft#1{\smash{\mathop{\longleftarrow}\limits^{#1}}}

\def\beq{\begin{equation}}
\def\eeq{\end{equation}}
\def\eqref#1{(\ref{#1})}

\def\symmref{EMS1,CGbook,KrV,Olv1,Olv2,Ste}

\def\solvref{BP,BH,HaA,MS,ShP}

\def\i{{(i)}}
\def\j{{(j)}}
\def\k{{(k)}}

\begin{document}

\title[Reduction of ODEs via $\s$-symmetries]{A generalization of $\lambda$-symmetry
reduction for systems of ODEs: $\s$-symmetries}

\author{G. Cicogna$^1$, G. Gaeta$^2$ and S. Walcher$^3$}

\address{{}$^1$ Dipartimento di Fisica, Universit\`a di Pisa, and INFN
sezione di Pisa, Largo B. Pontecorvo 3, 56127 Pisa (Italy); e-mail: {\tt
cicogna@df.unipi.it} \\
{}$^2$ Dipartimento di Matematica, Universit\`a degli Studi di Milano, via C. Saldini 50, 20133 Milano (Italy); e-mail: {\tt giuseppe.gaeta@unimi.it} \\
{}$^3$ Lehrstuhl A f\"ur Mathematik, RWTH Aachen,
52056 Aachen (Germany); e-mail: {\tt walcher@mathA.rtwh-aachen.de}}

\begin{abstract}
We consider a deformation of the prolongation operation, defined
on sets of vector fields and involving a mutual interaction in the
definition of prolonged ones. This maintains the ``invariants by
differentiation'' property, and can hence be used to reduce ODEs
satisfying suitable invariance conditions in a fully algorithmic
way, similarly to what happens for standard prolongations and
symmetries. 

\end{abstract}




\section*{Introduction}

Consider a (scalar) ODE $\E$ of order $n > 1$ in $J^n M$ (the Jet
bundle of order $n$ over the manifold $M$ of independent and
dependent variables); it is well known that if this admits a
(Lie-point) symmetry, then it can be reduced to an equation of
order $(n-1)$. More generally, if it admits a $p$-dimensional Lie
group of symmetries $G$, then it can be reduced to an equation of
order $(n-q)$, with $q \le p$ depending on the algebraic structure
of the operation of $G$ and its Lie algebra $\G$ \cite{\symmref}.

The approach to symmetry reduction for ODEs goes roughly speaking
as follows: if $\E$ admits the vector field $X$ (on $M$) as a
symmetry (generator), then $\E$ is invariant under $Y = X^{(n)}$,
the prolongation of $X$ to $J^n M$. Thus $\E$ can be expressed in
terms of the differential invariants of $Y$. Or, this can be
recursively generated starting from those of order zero and one
thanks to the ``invariants by differentiation'' property (see
\cite{\symmref}).

Using symmetry adapted coordinates, say $w$ and $y$ with $y$ the
independent variable,  the equation will not depend explicitly on
the coordinate along the vector field $X$, say $w$, and thus its
order can be reduced by one passing to consider it as an equation
for $z = w_y$ and its derivatives.

It was observed by Muriel and Romero back in 2001 that the same
scheme works if $Y$ is not the (standard) prolongation of $X$ but
instead some kind of ``twisted'' prolongation, depending on an
arbitrary smooth (by this we will always mean $\C^\infty$)
function $\la : J^1 M \to R$ and hence called
``$\la$-prolongation'' \cite{MuRom1,MuRom2,MuRom3}.

The key to this result is that $\la$-prolonged vector fields still
have the property of ``invariants by differentiation'', so the
scheme working in the case of invariance under standardly
prolonged vector fields can still be applied.

The same holds for systems of ODEs, except that here a symmetry
will lead to reduction by one of the order of one of the
equations; in the case of systems of $n$ first order equation, a
symmetry will lead to reduction to a system of $n-1$ equations.
For multi-generators symmetry group, as usual the attainable
reduction will depend on the algebraic structure of the operation of the group.

It was shown by Pucci and Saccomandi \cite{PuS} that for scalar
ODEs the $\la$-prolonged vector fields are essentially the only
ones sharing the ``invariants by differentiation'' property; they
also provided a neat geometrical understanding of this fact.

In the present note we want to generalize the Muriel-Romero
approach, and consider cases where the equation (or system) is
invariant under a set of vector fields $Y_\i$ defined on $J^n M$,
and obtained from vector fields in $M$ under a further modified
version of $\la$-prolongations. More precisely, in this case the
modified prolongation operation will not act on the single vector
field, but rather on the set of vector fields (that is,
prolongation of each of them will involve the other ones). This
``joint-$\la$'' prolongation will depend on a matrix $\s$ defined
by a set of smooth functions $\s_{ij}$ on $J^1 M$, and will be
therefore also denoted as $\s$-prolongation (in the same way as
$\la$-prolongations took their name from the function $\la$).

We will show that even in this case the standard approach to
reduction sketched above is still valid, and hence the equation
can be reduced. More precisely, we will show that the ``invariants
by differentiation'' property still holds. This will follow by
explicit algebraic computation, but later on we will also discuss
the geometrical meaning of our result, generalizing
the result by Pucci and Saccomandi \cite{PuS}.

Finally, we note that the same construction with ``joint
$\la$-prolongation'' is considered in a companion paper
\cite{CGW}; however there the assumptions, in particular on the
vector fields and the functions, are different: the two papers
are related, but deal with different situations and lead to
different results.

\bigskip\noindent
{\bf Acknowledgements.} This work was started in the course of a
visit of GG and SW in Pisa; they would like to thank the
Universit\`a di Pisa, and especially GC, for their hospitality in
that occasion.

\section{Preliminaries and notation}
\label{sec:prelim}

We will firstly recall some basic notion, also in order to set
some general notation to be used in the following.

\subsection{Equations, solutions, symmetries}

We will only consider ordinary differential equation(s); the
independent variable will be denoted as $x \in R$, the dependent
one(s) as $u \in U = R$ or $u^a \in U \sse R^p$ in the
multi-dimensional case. We denote by $M = X \x U$ the phase
bundle, and by $(J^k M , \pi^k , M)$, or $J^k M$ for short, the
associated jet bundle of order $k$.

A differential equation $\E$ of order $n$ is a map $F : J^n M \to
R$, and is naturally identified with the solution manifold $\S(\E)
= F^{-1} (0) \ss J^n M$. In the case of $\ell$-dimensional systems
$\E$, we have $\ell$ maps $F^j : J^n M \to R$ (or equivalently a
map $F : J^n M \to R^\ell$) and a solution manifold $\S(\E) =
F^{-1} ( {\bf 0} ) = (F^1)^{-1} (0) \cap ... \cap (F^\ell)^{-1}
(0) \ss J^k M$. Note that for $p$-dimensional dependent variables,
$J^n M$ has dimension $(n p + 1)$, and a system of $\ell$
independent equations identifies therefore a solution manifold of
dimension $(n p + 1 - \ell)$.

A function $f : X \to U$ is a solution to the differential
equation(s) $\E$ under study if and only if its $n$-th
prolongation lies entirely in the solution manifold $\S(\E)$.

Let us now consider a vector field $Y$ on $J^n M$; we say that
$\E$ is invariant under $Y$ if and only if its solution manifold
is; that is, $Y : \S(\E) \to \T \S (\E)$. This can also be cast as
the condition $ \[ Y ( \E ) \]_{\S(\E)} = 0$.

If $Y$ is the prolongation of a (Lie-point) vector field $X$ on
$M$, $Y = X^{(n)}$, we say that $X$ is a symmetry for $\E$ (more
precisely, this would be a symmetry generator; we will adopt this
standard abuse of notation for ease of language). The condition
for $X$ to be a symmetry is therefore
$ \[ X^{(n)} ( \E ) \]_{\S(\E)} = 0$.

\subsection{Local coordinates}

We will consider local coordinates $(x,u^a)$ in $M = X \x U$, and
correspondingly local coordinates $(x,u^a_k)$ (with $k=0,...,n$),
where $u^a_k := (\pa^k u^a / \pa x^k )$, in $J^n M$.

A general vector field on $J^n M$ will be written in local
coordinates (here and below we will use the Einstein summation
convention; the notation $u^{(k)}$ denotes $u$ together with its
derivatives of order up to $k$) as \beq\label{eq:Ygen} Y \ = \ \xi
(x,u) \, \frac{\pa}{\pa x} \ + \ \psi^a_k (x,u^{(k)} ) \,
\frac{\pa}{\pa u^a_k} \ ; \eeq this is the prolongation of
\beq\label{eq:Xgen} X \ = \ \xi (x,u) \, \frac{\pa}{\pa x} \ + \
\phi^a (x,u) \, \frac{\pa}{\pa u^a} \eeq if and only if the
coefficients $\phi^a_k$ satisfy the (standard) prolongation
formula \beq\label{eq:prolform} \psi^a_{k+1} \ = \ D_x \, \psi^a_k
\ - \ u^a_{k+1} \ D_x \xi \ \ ; \ \ \psi^a_0 = \phi^a \ . \eeq The
notation $D_x$ identifies the total derivative with respect to
$x$, \beq\label{eq:Dx} D_x \ = \ \pa_x \ + \ u^a_{k+1} \,
\frac{\pa}{\pa u^a_k} \ . \eeq

\subsection{$\la$-prolongations}

After the work of Muriel and Romero \cite{MuRom1,MuRom2} it is
frequent to also consider $\la$-symmetries of ODEs (see also
\cite{MuRom3,MuRom4,MuRom5,MuRom6,MRO} for their later work, and
the bibliography given in \cite{Gtwist}).

A vector field $Y$ written in local coordinates in the form
\eqref{eq:Ygen} is a $\la$-prolonged vector field if its
coefficients satisfy \beq\label{eq:laprol} \psi^a_{k+1} \ = \ D_x
\, \psi^a_k \ - \ u^a_{k+1} \ D_x \xi \ + \ \la \, \( \psi^a_k \ -
\ u^a_{k+1} \, \xi \) \eeq with $\la$ a smooth function $\la : J^1
M \to R$. We also say that $Y$ is the $n$-th $\la$-prolongation of
$X$ (see \eqref{eq:Xgen}) if $\psi^a_0 = \phi^a$, i.e. if $X$ is
the restriction of $Y$ to $M$.

If the $\la$-prolongation $Y$ of $X$ leaves an equation (or
system) $\E$ invariant, we say that $X$ is a $\la$-symmetry for
$\E$.

\subsection{The ``invariants by differentiation'' property}

When considering a vector field in $J^n M$, it is of interest to
know its invariants, i.e. the functions $\zeta : J^n M \to R$ such
that $Y(\zeta) = 0$; these are also called differential invariants
(to distinguish them from ``geometrical invariants'', i.e.
functions defined on $M$ rather than on $J^n M$).

For prolonged vector fields, it is well known that once we know
differential invariants of order zero and one, say $\eta$ and $\zeta_0$
respectively, we can generate recursively differential invariants
of any order. (Note that
starting from differential invariants of order one will also be
needed in the case of $\la$-prolongations in order to take into
account the properties of $\la$.) In fact, the functions \beq \zeta_{k+1} \ := \
\frac{D_x \zeta_k}{D_x \eta} \eeq turn out to be automatically
invariant under $Y$ if this is a prolongation and if $\eta$ and
$\zeta_k$ were \cite{\symmref}. This is also known as the {\it
invariants by differentiation property} (IBDP).

This property remains true in the case of vector fields which are
not standard prolongations but are $\la$-prolongations, and this
fact is at the basis of the approach by Muriel and Romero; see
e.g. their papers \cite{MuRom1,MuRom2} for an algebraic proof.

The validity of IBDP was understood in geometrical terms by Pucci
and Saccomandi \cite{PuS}; their argument can be recast in the
light of the approach by some of us \cite{CGM} as follows. Any
$\la$-prolonged vector field is collinear to a standardly prolonged vector field (the standard and the $\la$-prolongation being applied to different vector fields; the factor involved in the transformation can be nonlocal); however the invariants,
and hence also the IBDP, only detect the {\it direction} of vector
fields, not their magnitude (provided this is nonzero, of course),
and hence are the same for collinear vector fields.

In the following we will essentially generalize this remark to the
case where we have several vector fields, so that the scope for
variations in the generators of the resulting distribution is
ampler.

\section{Joint prolongations}
\label{sec:jointprol}

We consider a set ${\mathcal X} = \{ X_{(i)} , \ i = 1,...,r \}$
of vector fields $X_{(i)}$ in involution on $M$, i.e. such that
\beq\label{eq:invo} [ X_{(i)} , X_{(j)} ] \ = \ \mu_{ij}^k (x) \
X_k \eeq with $\mu_{ij}^k = - \mu_{ji}^k$ smooth functions on $M$.
We will also consider vector fields $Y_{(i)}$ on $J^k M$, which
will be some kind (to be specified in a moment) of  generalized
prolongation of the $X_{(i)}$.

In local coordinates, and with $u^a_k := (\pa^k u^a / \pa x^k)$,
these will be written as
\begin{eqnarray}
\label{eq:X} X_{(i)} &=& \xi_{(i)} (x,u) \, \frac{\pa}{\pa x} \ + \
\phi^a_{(i)} (x,u) \, \frac{\pa}{\pa u^a} \ , \\
\label{eq:Y} Y_{(i)} &=& \xi_{(i)} (x,u) \, \frac{\pa}{\pa x} \ +
\ \psi^a_{(i),k} (x,u^{(k)} ) \, \frac{\pa}{\pa u^a_k} \ ,
\end{eqnarray}
where we set $\psi^a_{(i),0} = \phi^a_{(i)}$.

In the following we will also use the shorthand notation \beq
\pa_a^k \ := \ (\pa / \pa u^a_k ) \ ; \eeq with this eq.
\eqref{eq:Y} reads \beq \label{eq:Ys} Y_{(i)} \ = \ \xi_{(i)} \,
\pa_x \ + \ \psi^a_{(i),k} \, \pa_a^k \ , \eeq

\medskip\noindent
{\bf Definition 1.} Let $\s = \{ \s_i^j \equiv \s_{ij} , \,
i,j=1,...,r \}$ be a set of $r^2$ smooth real functions on $J^1
M$. The vector fields $\mathcal{Y} = \{ Y_{(i)} , \, i=1,...,r \}$
on $J^k M$ ($i=1,...,r$), written in local coordinates as in
\eqref{eq:Y}, are said to be {\it jointly $\la$-prolonged} (or
{\it $\s$-prolonged}) if the coefficients $\psi^a_{(i),k}$
satisfy, for $k \ge 0$, \beq\label{eq:prol} \psi^a_{(i),k+1} =
\( D_x \psi^a_{(i),k} - u^a_{k+1} D_x \xi_{(i)} \) +
\s_{ij} \( \psi^a_{(j),k} - u^a_{k+1} \xi_{(j)} \) \ .
\eeq If $\psi^a_{(i),0} = \phi^a_{(i)}$, we say that the
$\mathcal{Y} = \{ Y_{(i)} \}$ are the {\it joint
$\la$-prolongation} (or {\it $\s$-prolongation} for short) of the
$\mathcal{X} = \{ X_{(i)} \}$.
\bigskip

In the following we use the notation with lower indices ($\s_{ij}$) for ease of writing when no confusion arises, resorting to the one with upper and lower ones ($\s_i^{\ j}$) when it becomes convenient to keep fully track of covariant and contravariant indices.

\medskip\noindent
{\bf Remark \ref{sec:jointprol}.1.} Note that the notion of
$\s$-prolongation refers to a {\it set of vector fields}, not to a
single one. \EOR

\medskip\noindent
{\bf Remark \ref{sec:jointprol}.2.} As $\s_{ij} : J^1 M \to \R$, this
notion represents indeed a generalization of $\la$-symmetries. At
the same time, the fact we introduce matrices rather than scalar
functions to describe the ``twisting'' of the prolongation
operation makes this similar to $\mu$-symmetries (in their ODEs version)
\cite{Gtwist}. Remark \ref{sec:prelim}.1 above guarantees we are
considering a really new notion, and we will see below it gives
new results. \EOR

\medskip\noindent
{\bf Lemma 1.} {\it If the $\Y$ are $\s$-prolonged, they satisfy
\beq\label{eq:mur0} [ Y_{(i)} , D_x ] \ = \ \s_{ij} \, Y_{(j)} \ - \ \( D_x
\xi_{(i)} \, + \, \s_{ij} \, \xi_{(j)} \) \, D_x  \ . \eeq}

\medskip\noindent
{\bf Proof.} This follows from explicit computation. In fact,
\beq D_x \ = \ \pa_x \ + \ u^a_{k+1} \, \pa_a^k \ ; \eeq
recalling \eqref{eq:Ys} we have immediately
\beq\label{eq:mur1} [Y_{(i)} , D_x ] \ = \ \( \psi^a_{(i),k+1} \ - \
D_x \psi^a_{(i),k} \) \, \pa_a^k \ - \ (D_x \xi_{(i)} ) \, \pa_x \ . \eeq
using \eqref{eq:prol}, the r.h.s. of this can be rewritten as
\begin{eqnarray*} & & \[ - u^a_{k+1} \, D_x \xi_\i \ + \ \s_{ij} \( \psi^a_{\j , k} \, - \, u^a_{k+1} \, \xi_\j \) \] \, \pa_a^k \ - \ (D_x \xi_\i ) \, \pa_x \\
 &=& - (D_x \xi_\i ) \, D_x \ + \ \s_{ij} \, \( Y_\i - \xi_\j \pa_x \) \ - \ \s_{ij} \, \xi_\j \, (D_x - \pa_x ) \\
  &=& - \, \[ (D_x \, \xi_\i) \ + \ \s_{ij} \, \xi_\j \] \, D_x \ + \ \s_{ij} \, Y_\j \ ; \end{eqnarray*}
this completes the proof. \EOP

\medskip\noindent
{\bf Remark \ref{sec:jointprol}.3} It follows easily from \eqref{eq:mur1} that, conversely, if the $Y_\i$ satisfy \eqref{eq:mur0}, then they satisfy \eqref{eq:prol}, i.e. are a $\s$-prolonged set. \EOR

\medskip\noindent
{\bf Theorem 1.} {\it Let $\Y$ be a set of $\s$-prolonged vector
fields, and let $\eta, \zeta$ be independent common differential invariants of
order $k$ for all of them. Then \beq \Theta \ := \ \frac{D_x
\zeta}{D_x \eta} \eeq is a common differential invariant of order
$k+1$ for all of them.}

\medskip\noindent
{\bf Proof.} It is obvious that $\Theta$ is of order $k+1$. To
show that it is invariant under any of the $Y_\i$, we just proceed
by straightforward computation; first of all, by
$$ Y_\i (\Theta) \ = \ \frac{[Y_\i (D_x \zeta)] \cdot (D_x \eta) \ - \
(D_x \zeta)  \cdot [Y_\i (D_x \eta)]}{(D_x \eta)^2} \ := \
\frac{\chi}{(D_x \eta)^2} $$ it is clear we just have to show that
the numerator $\chi$ vanishes. On the other hand, we have
(recalling that by assumption $Y_\i (\zeta) = Y_\i (\eta) = 0$,
and using Lemma 1 above)
\begin{eqnarray*}
\chi &=& [Y_\i (D_x \zeta) ] \cdot (D_x \eta) \ - \
(D_x \zeta) \cdot [Y_\i (D_x \eta)] \\
 &=& \( [D_x (Y_\i \zeta) ] \, + \,
 [Y_\i , D_x ] \zeta \) \cdot (D_x \eta) \\ & & \ - \
 (D_x \zeta) \cdot \( [D_x (Y_\i \eta)] \, + \,
 [Y_\i , D_x ] \eta \) \\
 &=& \( [Y_\i , D_x ] \zeta \) \cdot (D_x \eta) \ - \
 (D_x \zeta) \cdot \( [Y_\i , D_x ] \eta \) \\
 &=& - \, \( D_x \xi_\i + \s_{ij} \xi_\j \) \ (D_x \zeta) \ (D_x \eta) \\
 & & \ + \
 (D_x \zeta) \ (D_x \xi_\i + \s_{ij} \xi_\j ) \ (D_x \eta) \ = \ 0 \ . \end{eqnarray*}
This shows indeed $\chi = 0$ and hence the Theorem. \EOP

\medskip\noindent
{\bf Corollary 1.} {\it If a complete basis of (independent)
invariants of order zero and one for $\Y$ are known, we can
successively generate a basis for invariants of all orders.}

\medskip\noindent
{\bf Proof.} One should only check that the independence of the
$\eta_\a$ and $\zeta_\b$ (differential invariants of order zero
and $k$) implies independence of derived invariants. This only
concerns functional independence of the $\Theta_{\a \b} := (D_x
\zeta_\b / D_x \eta_\a)$, and is shown e.g. in Olver \cite{Olv1}.
Note here (and there) we are using the fact there is only one
independent variable. \EOP
\bigskip

Note that in Theorem 1 the $\s_{ij}$ are arbitrary (smooth)
functions. It is clear that the relations between the $X_\i$ will
not be shared by the $Y_\i$ for arbitrary $\s_{ij}$; the condition
under which the $Y_\i$ have the same commutation properties as the
$X_\i$ (i.e. the commutator of $\s$-prolonged vector fields in
$\Y$ is the same as the $\s$-prolongation of the commutator of
vector fields in $\X$), is discussed in Section \ref{sec:invol}
(under the simplifying assumptions that the vector fields do not
act on the independent variable).

\section{Involutivity of $\s$-prolonged sets of vector
fields} \label{sec:invol}

The IBDP for $\s$-prolonged vector fields, and hence the reduction
procedure to be described in the following, is based on the
condition that the $\s$-prolonged vector fields are in involution.
In this section we discuss conditions guaranteeing this is the
case, based on the preliminary condition that the $\{ X_\i \}$
are in involution.

\medskip\noindent
{\bf Theorem 2.} {\it Assume the vector fields $X_\i = \xi_\i
\pa_x + \phi^a_\i \pa_a$ ($i=1,...,n$) are in involution, with
$[X_\i , X_\j ] \ = \ \mu_{ij}^k \ X_{(k)}$ for $\mu_{ij}^k$ smooth functions on $M$. Then their $\s$-prolongations $Y_\i$ satisfy the same
involution relations, i.e. $ [Y_\i , Y_\j ] = \mu_{ij}^k Y_{(k)}$,
if and only if the $\s_i^{\ j}$ satisfy, for all $k=1,...,n$, the equations
\beq\label{eq:lagen} \begin{array}{l}
\{ [ Y_i (\s_j^{\ k} ) - Y_j (\s_i^{\ k}) ] \ + \\
\ \ + \ \( (D_x \mu_{ij}^k) + \s_i^{\ m} \mu_{mj}^k  -  \s_j^{\ m}
\mu_{mi}^k  - \mu_{ij}^m \s_m^{\ k} \) \} \ \phi^a_\k \ = \ 0 \ . \end{array}
\eeq}

\medskip\noindent
{\bf Proof.} This follows from an explicit computation, described
hereafter; we proceed by induction on the order of the
prolongation. Denoting by $Z_i$ the
$(q-1)$-th $\s$-prolongation of the $X_i$, we have
$$ [ Y_\i , Y_\j ] \ = \
[Z_\i , Z_\j ] \ + \ [ Y_\i (\psi^a_{\j,q} ) \, - \, Y_\j
(\psi^a_{\i,q} ) ] \ \pa_a^q \ := \ [Z_\i , Z_\j ] \ + \ F_{ij,q}^a
\, \pa_a^q \ . $$ Thus, assuming $[Z_\i , Z_\j ] = \mu_{ij}^k
Z_\k$ (i.e. the involution relations are satisfied for $(q-1)$-th
prolongations), the requirement that $[Y_\i , Y_\j ] = \mu_{ij}^k
Y_k$  is equivalent to the requirement that \beq\label{eq:Fcomm}
F^a_{ij,q} \ = \ \mu_{ij}^k \ \psi^a_{\k,q} \ . \eeq

The $F^a_{ij,q}$ needs to be rewritten for easier comparison with
the r.h.s. of \eqref{eq:Fcomm}. In the course of this computation
we will write, for a short notation,
$\Phi^a_i$ for $\psi^a_{\i,q-1}$; note this entails $Y_i
(\Phi^a_j) = Z_i (\Phi^a_j)$; with this, \eqref{eq:Fcomm} reads
\beq\label{eq:Fcomm2} F^a_{ij,q} \ = \ \mu_{ij}^k \ \( D_x \Phi^a_k
\, + \, \s_{km} \Phi^a_m \) \ . \eeq Using also Lemma 1, with
standard algebra we get (omitting the vector indices $a$ for ease
of writing -- and reading)
\begin{eqnarray*}
F_{ij,q} &=& Y_\i (\psi_{\j,q} ) \, - \, Y_\j (\psi_{\i,q}) \\
&=& Y_i [ (D_x \Phi_j - u_k D_x \xi_j ) + \s_{jk} (\Phi_k - u_k
\xi_k ) ] \\
& & \ - \ Y_j [ (D_x \Phi_i - u_k D_x \xi_i ) + \s_{ik}
(\Phi_k - u_k \xi_k ) ] \\
&=& (Y_i D_x \Phi_j - Y_j D_x \Phi_i ) \ + \ [ Y_i (\s_{jk}) - Y_j
( \s_{ik} ) ] \, (\Phi_k - u_k \xi_k ) \\
 & & \ - \ [\la_{ik} Y_j
(\Phi_k) - \la_{jk} Y_i (\Phi_k ) ] \\
&=& D_x [ Y_i (\Phi_j) - Y_j (\Phi_i ) ] \ + \ [ \s_{ik} Y_k
(\Phi_j) - \s_{jk} Y_k (\Phi_i) ] \\
 & & \ + \ [ Y_i (\s_{jk} ) - Y_j (\s_{ik}) ] \, \Phi_k \ - \ [
 \s_{ik} Y_j (\Phi_k) - \s_{jk} Y_i (\Phi_k ) ] \\
&=& D_x [ Z_i (\Phi_j) - Z_j (\Phi_i ) ] \ + \ [ \s_{ik} Z_k
(\Phi_j) - \s_{jk} Z_k (\Phi_i) ] \\
 & & \ + \ [ Y_i (\s_{jk} ) - Y_j (\s_{ik}) ] \, \Phi_k \ - \ [
 \s_{ik} Z_j (\Phi_k) - \s_{jk} Z_i (\Phi_k ) ] \\
 &=& D_x ( \mu_{ij}^k \Phi_k ) \ + \ [ Y_i (\s_{jk} ) - Y_j (\s_{ik}) ] \,
 \Phi_k \\
 & & \ + \ \s_{ik} [ Z_k (\Phi_j) - Z_j (\Phi_k ) ] \ - \
 \s_{jk} [ Z_k (\Phi_i) - Z_i (\Phi_k ) ] \\
 &=& (D_x  \mu_{ij}^k) \Phi_k \ + \ \mu_{ij}^k D_x (\Phi_k ) \ + \
 [ Y_i (\s_{jk} ) - Y_j (\s_{ik}) ] \, \Phi_k \\
 & & \ + \ \s_{ik} \, \mu_{kj}^\ell \Phi_\ell \ - \
 \s_{jk} \, \mu_{ki}^\ell \Phi_\ell \ .  \end{eqnarray*}

Comparing this with \eqref{eq:Fcomm2}, we must require
\begin{eqnarray*}
F_{ij} &:=& (D_x  \mu_{ij}^k) \Phi_k \ + \ \mu_{ij}^k D_x (\Phi_k
) \ + \
 [ Y_i (\s_{jk} ) - Y_j (\s_{ik}) ] \, \Phi_k \\
 & & + \s_{ik} \, \mu_{kj}^\ell \Phi_\ell \, - \,
 \s_{jk} \, \mu_{ki}^\ell \Phi_\ell \, = \, \mu_{ij}^k ( D_x \Phi_k)
\, + \, \mu_{ij}^k \s_{km} \Phi_m \, . \end{eqnarray*} That is,
eliminating equal terms on both sides and renaming the summation
indices,
\begin{eqnarray*}
& & (D_x  \mu_{ij}^k) \Phi_k \ + \
 [ Y_i (\s_{jk} ) - Y_j (\s_{ik}) ] \, \Phi_k \\
 & & \ + \ \s_{im} \, \mu_{mj}^k \Phi_k \ - \
 \s_{jm} \, \mu_{mi}^k \Phi_k
 \ - \ \mu_{ij}^m \s_{mk} \Phi_k \ = \ 0 \ . \end{eqnarray*}
We can now collect the $\Phi_k$ term, and finally get (reinserting the index $a$)
\beq\label{eq:Phi} \{ [ Y_i (\s_{jk} ) - Y_j (\s_{ik}) ] \, + \, (D_x  \mu_{ij}^k)
\, + \, \s_{im} \, \mu_{mj}^k \, - \,
 \s_{jm} \, \mu_{mi}^k
 \, - \, \mu_{ij}^m \, \s_{mk} \} \, \Phi^a_k \, = \, 0 \, . \eeq

We should now recall that $\Phi^a_k := \psi^a_{(k),q-1}$, and that \eqref{eq:Phi} was the condition under which the commutation properties holding up to order $q-1$ are also holding at order $q$.

On the other hand, since by Definition 1 the $\s_{ij}$ are functions on $J^1 M$, and $D_x \mu_{ij}^k$ are also functions on $J^1 M$ since $\mu_{ij}^k$ are functions on $M$, all terms within the curly brackets in \eqref{eq:Phi} only depends on $J^1 M$. That is, if \eqref{eq:Phi} is satisfied for first prolongations (i.e. for $q=1$), it will be automatically satisfied for all higher order prolongations (i.e. for $q > 1$) as well.

For $q=1$ we have $\Phi^a_j = \psi^a_{(j),q-1} = \phi^a_\j$, hence \eqref{eq:Phi} reduces precisely to \eqref{eq:lagen}, and the proof is now complete. \EOP

\medskip\noindent
{\bf Corollary 2.} {\it If, in the hypotheses and with the
notation of Theorem 2, it is \beq\label{eq:la} Y_i (\s_j^{\ k} ) -
Y_j (\s_i^{\ k})  = - \[ (D_x \mu_{ij}^k) + \s_i^{\ m} \mu_{mj}^k
-  \s_j^{\ m} \mu_{mi}^k - \mu_{ij}^m \s_m^{\ k} \] , \eeq then
the $\{ Y_\i \}$ satisfy the same involution relations as the $\{
X_\i \}$.}

\medskip\noindent
{\bf Proof.} In this case equation \eqref{eq:lagen} is automatically
satisfied. We stress \eqref{eq:la} is a sufficient but not
necessary condition for \eqref{eq:lagen} to hold. \EOP

\medskip\noindent
{\bf Remark \ref{sec:invol}.1.} The equation \eqref{eq:la}
actually involves only the first $\s$-prolongations, the $\s_{ij}$
being functions on $J^1 M$. Thus the r.h.s. of the relation \eqref{eq:la} is
rewritten as \begin{eqnarray}
& & \phi^a_\i \frac{\pa \s_j^{\
k}}{\pa u^a} \ + \ (D_x \phi^a_\i + \s_i^{\ \ell} \phi^a_{(\ell)}
) \frac{\pa \s_j^{\ k}}{\pa u^a_x} \ + \nonumber \\
& & \ \ - \ \phi^a_\j \frac{\pa
\s_i^{\ k}}{\pa u^a} \ + \ (D_x \phi^a_\j + \s_j^{\ \ell}
\phi^a_{(\ell)} ) \frac{\pa \s_i^{\ k}}{\pa u^a_x} \ . \label{eq:la3} \end{eqnarray}
Note that the \eqref{eq:la} are nonlinear in the $\s$; thus determining
suitable $\s$ preserving the involution properties of a given set
of vector fields (under $\s$-prolongation) is in general a
nontrivial task. \EOR

\medskip\noindent
{\bf Remark \ref{sec:invol}.2.} If the $X_\i$ commute, so that
$\mu_{ij}^k = 0$, then \eqref{eq:lagen} reduces simply to $[ Y_i
(\s_j^{\ k} ) - Y_j (\s_i^{\ k}) ] \ \Phi^a_{(k)} = 0$, and
\eqref{eq:la} to \beq  Y_i (\s_j^{\ k} ) - Y_j (\s_i^{\ k}) \ = \
0 \ . \eeq {} \EOR

\medskip\noindent
{\bf Remark \ref{sec:invol}.3.} We stress that the preservation of
involution relations is {\it not} required by the definition of
$\s$-prolonged sets of vector fields, nor it will be required for
the reduction of (systems of) differential equations based on
$\s$-symmetries. On the other hand, the $Y_\i$ should be an
involution system, and this is {\it not} guaranteed {\it apriori}
by the fact the $X_\i$ are in involution, unless equation
\eqref{eq:la}, or at least \eqref{eq:lagen}, is satisfied. \EOR

\section{Reduction and $\s$-symmetries}
\label{sec:reduction}

We will now consider the case where a given differential equation,
or rather system of differential equations, is invariant under a
set of $\s$-prolonged vector fields $\Y = \{ Y_{(1)} , ... ,
Y_{(N)} \}$. We will denote, as above, by $\X = \{ X_{(1)} , ... ,
X_{(N)} \}$ the set of vector fields in $M$ which, upon
$\s$-prolongation, yield the set $\Y$.

We will start by considering a (rather special) situation where
the construction needed for reduction is particularly transparent,
and then proceed to consider generalizations; we trust this way of
proceeding will help the reader.

\subsection{Full reduction}

Let us consider a system of ODEs $\E = \{ E^1 = 0 , ... , E^m = 0
\}$ of order $q$ for dependent variables $u = (u^1,...,u^n)$, i.e.
written as $F^h (x, u, ..., u^{(q)} ) =0$, where $h=1,...,m$.

We recall preliminarily that if $\E$ is invariant under a set $\Y$
of vector fields in $J^q M$, then it is equivalent to an equation
which can be written in terms of the joint differential invariants
(of order up to $q$) of $\Y$, see e.g. \cite{CDW}; we will thus
deal with equations written in this form. (One also says
that the equation admitting $\Y$ as a system of symmetries are
equivalent to equations admitting it as a system of strong
symmetries. We will systematically use the equivalent form, also
in Examples below.)

\medskip\noindent
{\bf Theorem 3.} {\it Let $\X$ be an involution system of rank
$n$ of vector fields over $M = \R \times \R^n$; and let $\Y$ be set, also of
rank $n$, of their $\s$-prolongations. Let the system of $m$
ordinary differential equations $\E (x,u^{(q)})$ of order $q > 1$ in
$n$ dependent variables be invariant under the set $\Y$, i.e.
admit $\X$ as $\s$-symmetries. Then $\E$ can be reduced to a
system of $m$ differential equations of order $q-1$.}

\medskip\noindent
{\bf Proof.} Let us consider the set $\X$ of vector fields in $M$
which, upon $\s$-prolonga\-ti\-on, yield the set $\Y$. We have
supposed the number of dependent variables is the same as the rank
of the set of vector fields in the set $\Y$, and therefore also in
the set $\X$. This means that the distribution $\D (\X)$ spanned
by the $\X$ in $\T M$ has dimension $n$, and hence that we can
introduce local coordinates $(y; w^1,...,w^n)$ in $M$ such that
$X_\i (y) = 0$ for all $i$. (Note this will in general mix $x$ and
the $u$, i.e. dependent and independent coordinates; this is
analogous to the standard situation \cite{Olv1,Ste}.)

Consider now the action of $\Y$ in $J^1 M$, which has dimension
$d_1 = (2n+1)$; again the distribution generated by the $\Y$ has
dimension $n$, and hence there exist $n$ differential invariants
of order one (in addition to $\eta = y $, invariant of order
zero). We denote these by $(\z_1 , ... \z_n )$.

When considering the action of $\Y$ in $J^q M$, we note the latter
has dimension $d_q = [(q+1) n + 1]$, while $\D (\Y)$ still has
dimension $n$, hence we have $(q n + 1)$ invariants; one of these
is $\eta$, and we have $n$ differential invariants of each order.

The latter can be built using Theorem 1; in the $(y;w)$
coordinates the higher order differential invariants $\z_a^{(k)}$
of order $k+1 \le q$ (thus $0 \le k \le  q-1$) will simply be
\beq\label{eq:thm3.1} \z_a^{(k)} \ = \ D_y^k \, \z_a \ , \ \ \ (a =
1 , ... , n ) \ . \eeq

As the system is supposed to be invariant under the $\Y$, the
differential equations can be written in terms of the common
differential invariants of $\Y$, i.e. in view of our discussion as
\beq\label{eq:thm3.2} F^h \( y ; \zeta , \zeta^{(1)} , ... ,
\zeta^{(q-1)} \) \ = \ 0 \ . \eeq

Introducing now new coordinates $z_i = \zeta_i$ ($i = 1, ... ,
n$), and recalling \eqref{eq:thm3.1} -- which implies that
$\z_a^{(k)} = D_y^k \z_a = D_y^k z_a$ -- the system
\eqref{eq:thm3.2} will be written as \beq\label{eq:thm3.3} F^h \( y
; z , z^{(1)} , ... , z^{(q-1)} \) \ = \ 0 \ , \eeq where of
course $ z^{(k)}$ denotes the set of $k$-th order derivatives
$(d^k z_i / d y^k)$.

We have thus constructively obtained the reduction of the initial
set of equations $F^h (x,u,...,u^{(q)} ) = 0$ to a set $\^\E$ of
equations \eqref{eq:thm3.3} of order $(q-1)$.

Finally we note that the requirement $q>1$ guarantees
that all invariants of order $q$  are built recursively from those of order
zero and one (see Corollary 1); we could not say the same for $q=1$ (as also obvious from the fact the action at order zero has no trace of the $\s_{ij}$ functions). \EOP

\medskip\noindent
{\bf Remark \ref{sec:reduction}.1.} Needless to say, the situation
we are considering in Theorem 3, i.e. as many symmetries (or more
precisely the rank of their system) as dependent variables in the
system, is highly special and rarely met in practice. As
anticipated at the beginning of the section, we are discussing
this case first as it helps in visualizing the mechanism under our
general reduction results, which will hold under much less
exacting conditions. The geometrical meaning of Theorem 3 is quite
transparent and shows easily the direction of generalization:
indeed we are using the fact the system $\Y$ on $J^q M$ admits an
invariant distribution (of dimension $n$), and hence common
integral manifolds, in $\T (J^q M)$, and use this information to
reduce the equations. The fact $n$ is precisely the number of
dependent variables allows to implement this reduction by lowering
the order of the full system, while in general we will have less
uniform reduction, as discussed below. \EOR

\medskip\noindent
{\bf Remark \ref{sec:reduction}.2.} Note that if we have solutions
of the reduced equations $\^\E$, i.e. explicit functions $z_i
(y)$, in order to have solutions of the initial equations $\E$ we
must solve the {\it reconstruction equations system}
\beq\label{eq:recon} z_i (y) \ = \ \z_i (y,w,w_y) \eeq as a
differential equations system for the $w(y) = \{ w_1 (y) , ... ,
w_n (y) \}$. After this, we will also have to invert the change of
coordinates $(x,u) \to (y,w)$ in order to express the solution in
the original coordinates in which the $\E$ where set. \EOR

\medskip\noindent
{\bf Remark \ref{sec:reduction}.3.} We stress that the reconstruction
equations \eqref{eq:recon} are a (generally) nonlinear system of non-autonomous equations. Thus, we are definitely not guaranteed to be able to solve them. \EOR

\medskip\noindent
{\bf Remark \ref{sec:reduction}.4.} As the differential invariants
thus built are common to all the vector fields, they are in
particular differential invariants for any one of them. Suppose
one of them, say $Y_{(0)}$ represents the differential equation we
want to reduce. The differential invariants can then be used to
perform the reduction along the lines of standard reduction for
ODEs with symmetry, see again Olver \cite{Olv1}. Note also that if
we wish $Y_{(0)}$ to be the standard prolongation of $X_{(0)}$, we
should just require that $\s_{0j} = 0$ for all $j$. \EOR

\subsection{Partial reduction}

In Theorem 3, the set $\mathcal{Y}$ was of the same rank $n$ as
the number of dependent variables. Needless to say, this is a very
special and fortunate case; in general, we will have (or be able
to determine) a set of rank $r < n$.

It should be clear by inspection that the approach followed in the
proof of Theorem 3 can be followed also in this case, leading of
course to somewhat different conclusions. We state this as a
variant to Theorem 3:

\medskip\noindent
{\bf Theorem 4.} {\it Let the system of $m$ ordinary differential
equations $\E (x,u^{(q)})$ in $n$ dependent variables be invariant
under the set $\Y$, of rank $r < n$, of $\s$-prolonged vector
fields. Then $\E$ can be reduced to a system of $m$ differential
equations, depending on derivatives of order up to $q>1$ for $(n-r)$
variables and on derivatives of order up to $(q-1)$ for $r$
variables. In particular, if the Jacobian $(\delta F / \delta
u^{(q)})$ is nonsingular, it can be reduced to a system of $m$
equations, $r$ of them of order $q-1$, and $(m-r)$ of them still
of order $q$.}

\medskip\noindent
{\bf Proof.} Proceeding as in the proof to Theorem 3, we determine
$r$ independent invariant functions $\zeta_i (x,u,u_x)$, $i =
1,...,r$; up to relabelling of variables we can then pass to new
variables $(v_i,z_j)$ defined by
$$ v_i = u_i \ \ \mathrm{for} \ 1 \le i \le s = (n-r) \ , \ \
z_j \ = \ \zeta_j \ \ \mathrm{for} \ 1 \le i \le r \ . $$ The
functions $F^h$ defining the ODEs under study can then be written
as
$$ F^h (x; v_1,...v_s; z_1,...,z_r; ... ; v_1^{(q-1)} , ... , v_s^{(q-1)}
; z_1^{(q-1)}, ... , z_n^{(q-1)}; v_1^{(q)},...,v_s^{(q)}) \ . $$
In other words the system is of order $q$ in the variables $v$ and
of order $(q-1)$ in the variables $z_j$; we can thus rewrite it as
in the statement. \EOP

\medskip\noindent
{\bf Remark \ref{sec:reduction}.5.} Needless to say, one could as
well relax the assumption that all equations are of the same
order, obtaining a corresponding special result by the same
procedure. The necessary modifications are clear and we leave details to
readers, possibly with a specific application in view. \EOR

\section{Geometrical features of $\s$-prolongations}
\label{sec:geo}

The result of
Theorem 1 can appear rather surprising: in fact, the ``invariants
by differentiation'' property is intimately related to the
standard prolongation structure, and one should think that if it
works for a somewhat arbitrary (or apparently so at first sight)
modified structure of vector fields there is some geometrical
reason. This is indeed the case, as we discuss in the present
section.

Let us consider a set of $\s$-prolonged vector fields $\{ Y_i \}$
in $J^n M$. As remarked before (see Section \ref{sec:prelim})
following Pucci and Saccomandi, the invariants only depend on the
distribution spanned by these vector fields; in other words,
passing to a different set of vector fields $\{ Z_i \}$ which are
point-wise in the linear span of the $Y_i$, we have the same
distribution and hence in particular the same set of invariant
functions in $J^n M$. That is invariants are related to the module
(over $\mathcal{C}^\infty (J^n M , R)$ function) generated by $\{
Y_i \}$, and are the same for any choice of the module generators.

We want to discuss if the same module with $\s$-prolonged
generators $\{ Y_i \}$ also admits generators $\{ Z_i \}$ which
are {\it standardly prolonged} vector fields. We will look for
these in the form \beq\label{eq:ZY} Z_i \ = \ A_{ij} \ Y_j \ ,
\eeq where the $A_{ij}$ are smooth functions on $M$, and $A$ is
non-singular at all points; note that as $A$ is a function on $M$,
the restriction of $Z_i$ to $J^h M$ is a proper vector field on
$J^h M$ (since this is the case for $Y_i$). It clearly suffices to
discuss the situation for first prolongations, as $k$-th
prolongations are obtained as first prolongations of $(k-1)$-th
prolongations.

\medskip\noindent
{\bf Theorem 5.} {\it Let $\{ X_i \}$ be vector fields in $M$, and
$\{ Y_i \}$ be their $\s$-prolongations. The module generated by
the set of $\s$-prolonged vector fields $Y_i$ obtained for $\s =
\s_{ij}$ coincides with the module generated by the standard
prolongations $Z_i$ of the vector fields $W_i = A_{ij} X_j $ with
$A_{ij}$ smooth functions on $M$ satisfying \beq\label{eq:A} D_x
\, A \ = \ A \ \s \ . \eeq}

\medskip\noindent
{\bf Proof.} The argument is specially transparent if we deal with
vertical vector fields in $M$, i.e. with vector fields acting only
on dependent variables, and we will start by considering this
case. We will work in local coordinates, and write
\beq\label{eq:XY} X_i \ = \ \phi^a_i \, \frac{\pa}{\pa u^a} \ ; \
\ Y_i \ = \ \phi^a_i \, \frac{\pa}{\pa u^a} \ + \ \( D_x \phi^a_i
\, + \, \s_{ij} \phi^a_j \) \, \frac{\pa}{\pa u^a_x} \ . \eeq

We look for vector fields $W_i$ of the form \beq\label{eq:W} W_i \
= \ A_{ij} \ X_i \ = \ \( A_{ij} \, \phi^a_j \) \ \frac{\pa}{\pa
u^a} \ ; \eeq the standard prolongation of these will be
\beq\label{eq:Z} Z_i \ = \ \( A_{ij} \, \phi^a_j \) \
\frac{\pa}{\pa u^a} \ + \ \[ (D_x A_{ij} ) \, \phi^a_j \ + \
A_{ij} \, (D_x \phi^a_j) \] \ \frac{\pa}{\pa u^a_x} \ . \eeq In
view of \eqref{eq:XY}, \eqref{eq:W} and \eqref{eq:Z}, it is clear
that \eqref{eq:ZY} is satisfied if and only if \beq  (D_x A_{ij} )
\, \phi^a_j \ + \ A_{ij} \, (D_x \phi^a_j) \ = \ A_{ij} \ \( D_x
\phi^a_j \, + \, \s_{jk} \phi^a_k \) \ ; \eeq this is readily
simplified getting rid of the $A D_x \phi$ factors, yielding
(after rearrangement of summation indices) \beq  (D_x A_{ik} ) \,
\phi^a_k  \ = \ A_{ij} \, \s_{jk} \phi^a_k  \ . \eeq We therefore
get that this is true for any choice of the $\phi^a_k$ provided
the matrix function $A$ satisfies \eqref{eq:A}.
\medskip

The computation is just slightly more involved in the case of
general vector fields. In this case \eqref{eq:XY} is replaced by
\beq\label{eq:XYgen} X_i = \xi_i \pa_x + \phi^a_i \pa_a \ , \ \
Y_i = X_i + \psi^a_i \pa_a^x \eeq with $$ \psi^a_i \ = \ (D_x
\phi^a_i - u^a_x D_x \xi_i ) \ + \ \s_{ij} (\phi^a_j - u^a_x \xi_j
) \ ; $$ and $$ W_i = \chi_i \pa_x + \eta^a_i \pa_a \ , \ \ Z_i
= W_i + \Theta^a_i \pa_a^x $$ where (standard prolongation formula)
$$ \Theta^a_i \ = \ D_x \eta^a_i - u^a_x D_x \chi_i \ . $$
Setting $W_i = A_{ij} X_j$ yields
$$ \chi_i = A_{ij} \xi_j \ , \ \ \eta^a_i = A_{ij} \phi^a_j \ . $$
Now, requiring that $Z_i = A_{ij} Y_j$ amounts to requiring that
$\Theta^a_i = A_{ij} \psi^a_j$; in view of the above formulas,
this is written as
$$ A_{ij} (D_x \phi^a_j - u^a_x D_x \xi_j ) \ + \
(D_x A_{ik}) (\phi^a_k - u^a_x \xi_k) \ = \ A_{ij} [ (D_x \phi^a_j
- u^a_x \xi_j) \ + \ \s_{jk} (\phi^a_k - u^a_x \xi_k ) ] \ ; $$
eliminating the equal terms on both sides, this yields
$$ ( D_x A_{ik} \, - \, A_{ij} \, \s_{jk} ) \
(\phi^a_k \, - \, u^a_x \, \xi_k ) \ = \ 0 \ . $$ Again, this
holds for any $(\phi^a_k, \xi_k)$ provided $A$ satisfies
\eqref{eq:A}. \EOP

\medskip\noindent
{\bf Remark \ref{sec:geo}.1.} As $A$ is invertible we can also
write the relation \eqref{eq:A} in the form \beq\label{eq:A2} \s \
= \ A^{-1} \, D_x A \ . \eeq Note that the solution to equation
\eqref{eq:A}, equivalently to \eqref{eq:A2}, will in general {\it
not} be unique; see Example 4. \EOR

\medskip\noindent
{\bf Remark \ref{sec:geo}.2.} It should be stressed that, as also appearing from \eqref{eq:A2}, the function $A$ can be a nonlocal one; in particular,  \beq\label{eq:A3} A \ = \ \exp \[ \int \s \, d x \] \eeq and unless $\s = D_x S (x,u)$ for some local function $S$ we get a nonlocal function. \EOR

\medskip\noindent
{\bf Remark \ref{sec:geo}.3.} It should be stressed that the set
$\{ Z_i \}$ and the set $\{ Y_i \}$ will in general {\it not} have
the same invariants. Moreover, they could have different
involution properties, see Example 5. \EOR
\bigskip

Theorem 5 has a converse, which we state as a Lemma (Lemma 2) in
view of its interest to build concrete examples. We also stress
that this construction, if applied to vector fields which are in
involution, will produce vector fields which are again in
involution (Lemma 3).

\medskip\noindent
{\bf Lemma 2.} {\it Let $\{ W_i \}$ be a set of vector fields on
$M$, and $\{ Z_i \}$ their standard prolongations. Consider linear
combinations of these with a nonsingular matrix function $A : M
\to \mathrm{Mat}(n)$, $Y_i = A_{ij} Z_j$. Then the $\{ Y_i \}$ are
a set of $\s$-prolonged vector fields with $\s$-prolongation
coefficients $\s_{ij} = A^{-1} (D_x A)$.}

\medskip\noindent
{\bf Proof.} Obvious by construction. \EOP

\medskip\noindent
{\bf Lemma 3.} {\it Let $\{ Y_i \}$ be a set of vector fields on
$J^k M$, and assume they are in involution, so that $[Y_i , Y_j ]
= \mu_{ij}^k Y_k$ for some functions $\mu_{ij}^k : J^k M \to \R$.
Consider linear combinations of these with a nonsingular matrix
function $A : M \to \mathrm{Mat}(n)$, $Z_i = A_{ij} Y_j$. Then the
$\{ Z_i \}$ are in involution, $[Z_i , Z_j ] = \theta_{ij}^k
Z_k$, with $\theta_{ij}^k : J^k M \to \R$ explicitly given by
$\ \theta_{ij}^k = \[ A_{im} \mu_{m \ell}^h A_{j \ell} + A_{im} Y_m (A_{jh} ) - A_{jm} Y_m (A_{ih}) \]  A^{-1}_{hk}$.}

\medskip\noindent
{\bf Proof.} For $Z_i$ as in the statement, it follows by standard
algebra that \begin{eqnarray*} [Z_i , Z_j] &=& A_{im} A_{jk}
[Y_m , Y_k] + [(A_{im} Y_m (A_{jk})] Y_k  -  [A_{jk}
 Y_m (A_{im})]  Y_m \\
&=& [A_{im} A_{j k} \mu_{mk}^\ell ] Y_\ell + [(A_{im}  Y_m
(A_{jk})]  Y_k  -  [A_{jk}  Y_m (A_{im})]  Y_m \\
&=& \[ \( A_{im} \mu_{m \ell}^h A_{j \ell} + A_{im} Y_m
(A_{jh} )  -  A_{jm} Y_m (A_{ih}) \) A^{-1}_{hk} \]  Z_k \ .
\end{eqnarray*} This proves the $Z_i$ are in involution, and moreover
provides the explicit form for the functions $\theta_{ij}^k$ in
terms of the functions $\mu_{ij}^k$ and of the transformation
matrix $A$, invertible by assumption. \EOP

\medskip\noindent
{\bf Remark \ref{sec:geo}.4.} Theorem 5 and its converse (Lemma 2)
can also be stated as the commutativity of the diagram
$$ \matrix{ \mathcal{W} & \mapright{\mathtt{prol}} & \mathcal{Z} \cr
 & & \cr
\mapdown{A} & & \mapdown{A} \cr
 & & \cr
\mathcal{X} & \mapright{\s-\mathtt{prol}} & \mathcal{Y} \cr}  $$
(where of course $\mathcal{X} = \{ X_i \}$, $\mathcal{Y} = \{ Y_i
\}$, and so on) if $A$ and $\s$ are related by $$ \s \ = \ A \ \(
D_x A^{-1} \) \ = \ - \, (D_x A) \, A^{-1} \ . $$
{} \EOR

\medskip\noindent
{\bf Remark \ref{sec:geo}.5.} Suppose we operate with a second
(invertible) linear transformation $B$, so that the vector fields
$X_i = \phi^a_i \pa_a$ and $Y_i = X_i + \psi^a_i \pa_a^x$ are
mapped into $\^X_i = B_i^{\ j} X_j = \^\phi_i^a \pa_a$ and $\^Y_i
= B_i^{\ j} Y_j = \^X_i + \^\psi_i^a \pa_a^x$. Then
$$ \^\psi_i \ = \ B_i^{\ j} \psi_j \ = \ B_i^{\ j} (D_x \phi_j +
\s_j^{\ k} \phi_k ) \ ; $$ writing $\phi = B^{-1} \^\phi$, we
obtain readily that the vector fields $\{ Y_i \}$ are the
$\^\s$-prolongation of the $\{ X_i \}$ (that is, satisfy $\^\psi_i
= D_x \^\phi_i + \^\s_i^{\ j} \^\phi_j$) with
$$ \^\s \ = \ B \, \s \, B^{-1} \ + \ B \, (D_x B^{-1} ) \ . $$
Note this reproduces \eqref{eq:A2} if we start from $\s = 0$.
Needless to say, this states that $\s$ transform as gauge
coefficients \cite{Gtwist}. \EOR

\medskip\noindent
{\bf Remark \ref{sec:geo}.6.} At first sight one could think that
Theorem 5 and Lemma 2 state the triviality of $\s$-prolongations,
in the sense that sets of $\s$-prolonged vector fields can be
mapped into (and hence are equivalent to) sets of standardly
prolonged ones. However, a little reflection shows that this
equivalence is in general only local; in particular if the
distribution spanned by the $Y_i$ is singular at some points (we
recall this can mean either that some vector fields become
singular, or that the rank becomes smaller, at specific points) so
that the domain on which our procedures are well defined is not
simply connected, then some cohomological effects will appear, and
the equivalence will hold only on contractible subsets of $M$ and
$J^k M$. The situation here is quite similar to the one met in
discussing $\mu$-prolongations \cite{CGM} (for other similarities
-- and differences -- between $\s$ and $\mu$-prolongations, see
the Appendix A). \EOR

\medskip\noindent
{\bf Remark \ref{sec:geo}.7.} It should be stressed again that the
results of this section are a direct generalization of those
obtained by Pucci and Saccomandi \cite{PuS} for standard
$\la$-prolongations. In fact, they showed that -- in the scalar
ODE case -- the most general class of vector fields in $J^k M$
having the same characteristics as a standardly prolonged one (and
as a consequence, sharing with them the IBDP) is precisely that of
$\la$-prolonged vector fields (except for a degenerate case,
$\rho_1 = \rho_2 = 0$ in their classification, corresponding to
contact symmetries). Here we have shown that -- in the case of
vector ODEs -- the most general class of sets of vector fields in
$J^k M$ having the same integral manifolds as a set of standardly
prolonged ones (and as a consequence, sharing with them the IBDP)
is precisely that of $\s$-prolonged sets of vector fields. If we
agree that our technique provides the proper generalization of
their ``telescopic vector fields'' (which were defined as those
having the same integral lines as some standardly prolonged vector
field) to the case where one has a system vector fields in
involution (rather than single ones) and correspondingly looks at
integral manifolds (rather than integral lines), our discussion
confirms their statement that ``telescopic vector fields seem to
be the natural framework for the study of reduction methods based
on differential invariants'' (see \cite{PuS}, p.6151). \EOR

\section{Discussion and possible generalizations}
\label{sec:discussion}

We have introduced a new modified prolongation
operator, called $\s$-prolongation or joint $\la$-prolongation;
this does not apply to vector fields individually, but instead to
a given set of vector fields in involution. We have shown that the
$\s$-prolonged vector fields still possess the ``invariant by
differentiation property'', and hence if they have suitable
relations with the vector field describing a system of ODEs they
can be used for reducing the ODE system.

We have also discussed the geometrical meaning of $\s$-prolonged
vector fields; we found that these generalize to higher dimensions
the property of $\la$-pro\-lon\-ga\-ti\-ons -- discovered by Pucci
and Saccomandi \cite{PuS} -- of having the same integral lines as
some other standardly prolonged vector field, and being the most
general vector fields in $J^k M$ (projectable to $J^h M$ for all
$0 \le h \le m$) with this property; needless to say, as we are in
higher dimension integral lines are here replaced by integral
manifolds.

Sets of vector fields which, after being $\s$-prolonged (with a
specific matrix $\s$), leave invariant a given system of ODEs have
been christened $\s$-symmetries. In general, vector fields which are $\s$-symmetries of a system are not ordinary symmetries as well; thus
our construction really gives new weapons to the arsenal of useful
and structurally interesting procedures for studying nonlinear systems.

As already mentioned, the same procedure is studied from a slightly
different perspective and with some difference in a companion paper \cite{CGW}.
In particular, in that paper one is allowing vector fields which are not symmetries of the evolution equation but which generate an involution system with the vector field representing it; and one also allows vector fields in $J^k M$ which are not
necessarily prolongations of vector fields in $M$, in the same way as the functions $\s_{ij}$ are not necessarily depending only on variables in $J^1 M$. From the present point of view, these represent {\it generalized} $\s$-symmetries (in the same sense as one usually speaks of generalized symmetries in standard symmetry analysis \cite{\symmref}); discussing these would be outside the limits of the present work.

The careful reader has probably noted that throughout the paper we discussed how to use $\s$-symmetries if we know them for a given equation, but avoided to discuss how to determine these. The reason for this is that the determining equations are in this case (as is also the case for $\lambda$-symmetries) functional equations -- and in this case matrix functional equations; fully solving them is thus in general far beyond reach, and one has to rely on educated guesswork (or sheer luck) in order to determine convenient special solutions (see also Appendix B in this respect), i.e. assume a given functional form for the $\s$. Luckily, one does not need to know the most general $\s$-symmetries to be able to use them, and special solutions can be enough to reduce the equations under study (this feature is again in common with $\la$-symmetries).

Note that on the other hand, determining the equations or systems which admit a given set of vector fields as a $\s$-symmetry (with assigned $\s$) is a more tractable problem; actually, once we have determined differential invariants of order zero and one, the IBDP makes it immediate to determine differential invariants of higher orders and hence invariants equations and systems.

Some other directions of further research can be foreseen, and we very
sketchily discuss them before passing to discuss a number of Examples illustrating our results; these are left for future investigations, by ourselves or by some readers of this note.

\begin{itemize}

\item[(1)] Here we mainly discussed the general case, i.e.
equations of arbitrary order $q>1$ with $n$ dependent variables and an
involution system of rank $d \le n$. The case of {\it dynamical
systems}, i.e. $q=1$, would be of obvious interest, and deserves
further investigation along the lines of the present paper; as already
mentioned, it is considered from a slightly different point of view
in the companion paper \cite{CGW}. A forthcoming paper
\cite{CGW2} contains a study along the lines of the present approach.

\item[(2)] Due to obvious physical reasons, one is specially
interested in systems of ODEs which arise from a {\it variational
principle} -- i.e. systems of Euler-Lagrange equations. These will
present special features, and it is an easy guess that
$\s$-symmetries will be specially effective in symmetry reducing
this kind of systems, exactly as it happens for standard
symmetries (Noether theorem) and for $\la$- and $\mu$-symmetries
\cite{CGN,MRO}.

\item[(3)] Lambda-symmetries are naturally related to {\it nonlocal}
(standard) symmetries \cite{CF,NuL,MuRom4}. It is natural to expect that, as $\s$-symmetries are a generalization of $\la$-symmetries, some relations exists between $\s$-symmetries and some type of nonlocal symmetries (possibly based on the construction of Section \ref{sec:geo}).

\item[(4)] Finally, we would like to mention that an extension of
the classical symmetry reduction is based on so called {\it solvable
structures}, see e.g. \cite{\solvref} for details. We expect a
corresponding generalization, along the lines discussed here,
would be possible; and actually quite natural as that theory is
naturally set in terms of distributions of vector fields rather
than of single ones. In this respect it should be noted that
already Pucci and Saccomandi suggested that ``a more geometric
theory of telescopic vector fields and $\la$-symmetries is surely
possible by means of the theory of solvable structures or the
theory of coverings'' (see \cite{PuS}, p.6154), so we are again
suggesting a generalization of their approach.

\end{itemize}

\section{Examples}

\subsection*{Example 1.}

Let us consider $X = R$ with coordinate $x$, $U = R^2$ with
coordinates $(u,v)$. In $M = X \times U$
the system \beq\label{eq:exa1.1} \cases{u_{xx} \ = \ u_x v_x (1 + e^{- u}) &
, \cr v_{xx} \ = \ u_x v_x (1 - e^{-v}) &  \cr} \eeq
is $\sigma$-symmetric under   the vector
fields
$$ X_1 \ = \ \pa_u \ , \ \ X_2 \ = \ \pa_v  $$
with the functions $\s_{ij}$ given in matrix form by
$$ \s \ = \ \pmatrix{0 & v_x \cr u_x & 0 \cr} \ . $$
The $\s$-prolonged vector fields $Y_i$ in $J^2 M$ are then given by
\begin{eqnarray*}
Y_1 &=& \pa_u \, + \, v_x \, \pa_{v_x} \, + \,
u_x v_x \, \pa_{u_{xx}} \, + \, v_{xx} \, \pa_{v_{xx}} \ ; \\
Y_2 &=& \pa_v \, + \, u_x \, \pa_{u_x} \, +
\, u_{xx} \, \pa_{u_{xx}} \, + \, u_x v_x \,
\pa_{v_{xx}} \ . \end{eqnarray*}
Note these are in
involution, and actually commute, $[Y_1,Y_2] = 0$. Note also, in
passing, that in this case equation \eqref{eq:la} is satisfied, as
easily checked; thus we are in the situation claimed by Theorem 2.

The only common geometrical invariant (differential invariant of
order zero) for these vector fields is obviously $\eta = t$;
common differential invariants of order one for $\Y = \{Y_1 , Y_2
\}$ are provided by
$$ \zeta^{(0)}_1 \ = \ e^{- u} \ v_x \ , \ \
\zeta^{(0)}_2 \ = \ e^{- v} \ u_x \ . $$

According to our Theorem 1, $\zeta^{(1)}_i = D_x [\zeta^{(0)}_i]$
should be differential invariants of order two. In fact, we have
$$ \begin{array}{l}
\zeta^{(1)}_1 \ = \ D_x \[ \zeta^{(0)}_1 \] \ = \
e^{- u} \ \( v_{xx} \, - \, u_x \, v_x \) \ , \\
{} \\
\zeta^{(1)}_2 \ = \ D_x \[ \zeta^{(0)}_2 \] \ = \ e^{- v} \
\( u_{xx} \, - \, u_x \, v_x \)  \ ;
\end{array} $$
it is immediate to check that these are indeed invariant under
both $Y_1$ and $Y_2$, as claimed by Theorem 1.

Let us now consider, to give an illustration of Theorem 3, any
second order differential equation(s) invariant under $\Y$; these
are necessarily written in the form
$$ F^h (t,z_1,z_2,w_1,w_2) \ = \ 0 \ , $$
where we have set, for ease of writing,
$$ z_1 = \zeta^{(0)}_1 \ , \ \ z_2 = \zeta^{(0)}_2 \ ; \ \
w_1 = \zeta^{(1)}_1 \ , \ \ w_2 = \zeta^{(1)}_2 \ . $$

The above system corresponds to the choice
$$ F^1 = w_1 + z_1 z_2 \ , \ \ F^2 = w_2 - z_1 z_2\ .$$
Through the change of coordinates
$$ z_1 \ = \ e^{- u} \, v_x \ , \ \ z_2 \ = \ e^{- v} \, u_x  $$
system \eqref{eq:exa1.1} will  reduce, eliminating
the common exponential factors, to the first order system
\beq \begin{array}{l}
d z_1 / d x \ = \ - \, z_1 \, z_2 \ , \\
d z_2 / d x \ = \ z_1 \, z_2 \ . \end{array} \eeq

\subsection*{Example 2.}

Let us again consider $X = R$ with coordinate $x$ and $U = R^2$ with
coordinates $(u,v)$. We take now the vector fields
$$ X_1 \ = \ u \, \pa_u \ , \ \ X_2 \ = \ - u \, \pa_v \ ; $$
these are in involution, and actually satisfy
$ [X_1 , X_2 ] = X_2$.

We consider the functions $\s_{ij}$ given in matrix form by
$$ \s \ = \ \pmatrix{0 & u \cr u_x & 0 \cr} \ . $$
The $\s$-prolonged vector fields $Y_i$ in $J^2 M$ are then given by
\begin{eqnarray*}
Y_1 &=& u \, \pa_u \, + \, u_x \, \pa_{u_x}
\, + \, u^2 \, \pa_{v_x} \\
& & \, + \, (u^2 u_x + u_{xx}) \, \pa_{u_{xx}}
\, + \, (3 u u_x) \, \pa_{v_{xx}} \ ; \\
Y_2 &=& - u\, \pa_v \, - \, u u_x \, \pa_{u_x} \, + \,
u_x \, \pa_{v_x} \\ & & \, - \, (u u_{xx} + 2
u_x^2) \, \pa_{u_{xx}} \, - \, (u_{xx} + u^2 u_x) \,
\pa_{v_{xx}} \ . \end{eqnarray*} Note these are in
involution, and actually satisfy $ [Y_1,Y_2] = Y_2$;
these are the same involution relations satisfied by $X_1$ and
$X_2$, and also in this case one can check that equation
\eqref{eq:la} is indeed satisfied (and thus Theorem 2 holds).

The only common geometrical invariant (differential invariant of
order zero) for $\X = \{X_1 , X_2 \}$ is obviously $\eta = x$;
common differential invariants of order one for $\Y = \{Y_1 , Y_2
\}$ are provided by
$$ \zeta^{(0)}_1 \ = \ e^{- v} \, \frac{u_x}{u} \ , \ \
\zeta^{(0)}_2 \ = \ 2 \, v_x \, - \, u^2 \, - \,
2 \, ( 1 - e^{- v}) \, \frac{u_x}{u} \ . $$

According to our Theorem 1, $\zeta^{(1)}_i = D_x [\zeta^{(0)}_i
]$ should be differential invariants of order two. In fact, we
have
\begin{eqnarray*}
\zeta^{(1)}_1 \ = \ D_x \[ \zeta^{(0)}_1 \] &=&
(e^{- v} / u^2) \ \( u u_{xx} - u_x^2 - u u_x v_x \)  \ , \\
\zeta^{(1)}_2  \ = \  D_x \[ \zeta^{(0)}_2 \] &=&
(2/x^2) \ \[ \( u_x^2 - u^3 u_x - u u_{xx} + u^2 v_{xx} \) \right. \\
& & \left. \, + \, \( u u_{xx} - u_x^2 - u u_x v_x \) \] \ . \end{eqnarray*}
It is immediate to check that these are indeed invariant under both $Y_1$ and $Y_2$.

Let us now consider, with a view at Theorem 3, the second order equations
\begin{eqnarray}
u_{xx} &=& (1 / u) \ \[ (4 e^{-v} - 7) u_x^2 + (9 u v_x - 4 u^3) u_x \right. \nonumber \\
 & & \left. \ + \
e^v \( u^6 + 4 (u_x^2 + u^2 v_x^2 + u^3 u_x - u^4 v_x - 2 u u_x v_x) \) \]  \ , \nonumber \\
v_{xx} &=& (1/u^2) \ \[ u^3 u_x + (e^{- 2 v} - 1) u_x^2 + u u_{xx} \right. \nonumber \\
 & & \left. \ - \ e^{-v} \( u u_{xx} - 2 u u_x v_x + (1/2) u^3 u_x \) \] \ . \label{eq:exa2.1} \end{eqnarray}

These are invariant under both $Y_1$ and $Y_2$, as can be checked
by explicit computations; passing to the $Y$-adapted coordinates
$\z_1,\z_2$ requires to determine the inverse change of
coordinates, which is (writing $w_i = d \z_i / d x$)
$$ \begin{array}{ll}
u_x \ = \ u \z_1 \ , \ \ &
u_{xx} \ = \ (u/2) [2 e^v w_1 + \z_1 (u^2 + 2 \z_1 + \z_2 )] \ , \\
v_x \ = \ (u^2 + \z_2 ) / 2 \ , \ \ & v_{xx} \ = \ (1/2) [ 2 (e^v
- 1) w_1 + w_2 + (3 u^2 + \z_2 ) \z_1 ] \ . \end{array} $$
Plugging these into the equations \eqref{eq:exa2.1}, these reduce
to \beq w_1 := d \z_1 / d x \ = \ \z_2^2 \ , \ \ w_2 :=
d \z_2 / d x \ = \ \z_1 \, \z_2 \ , \eeq i.e. to a system of
first order equations.

\subsection*{Example 3.}

In the previous two examples we have just remarked equation
\eqref{eq:la} was satisfied; in order to show this is not always
the case, and that failing to satisfy this equations leads in
general to $\s$-prolonged vector fields which are not in
involution, we consider again the vector fields of Examples 1 and
2 but with different matrices $\s$.

\subsubsection*{Case 1.}

In the case of Example 1, i.e. $X_1 = \pa_u$ and $X_2 = \pa_v$,
and a matrix $\s$ which is just the transpose of that used in
there, i.e.
$$ \s \ = \ \pmatrix{ 0 & u_x \cr v_x & 0 \cr } \ . $$
In this case the first joint prolongation is given by $Y_1 = \pa_u
+ u_x \pa_{v_x}$, $Y_2 = \pa_v + v_x \pa_{u_x}$; thus the
involution relations between the $X$ are not mapped into
corresponding relations between the $Y$; more precisely, $ [Y_1 ,
Y_2 ] = Y_3 := u_x \pa_{u_x} - v_x \pa_{v_x}$. Moreover, in order
to close the involution relations we also have to introduce two
other new vector fields, $ Y_4 := u_x \pa_{v_x}$ and $Y_5 := v_x
\pa_{u_x}$. With these, the involution (actually, algebraic -- due
to the constant coefficients in $X_1$ and $X_2$) relations are
then

\smallskip
\begin{tabular}{llll}
 $[ Y_1 , Y_2 ] = Y_3$ , & $[ Y_1 , Y_3 ] = - 2 \, Y_4$ , &
 $[ Y_1 , Y_4 ] = 0$ , & $[ Y_1 , Y_5 ] = Y_3$ , \\
 $[ Y_2 , Y_3 ] = 2 \, Y_5$ , & $[ Y_2 , Y_4 ] = - Y_3$ , &
 $[ Y_2 , Y_5 ] = 0$ , & {} \\
 $[ Y_3 , Y_4 ] = 2 \, Y_4$ , & $[ Y_3 , Y_5 ] = - 2 \, Y_5$ , &
 $[ Y_4 , Y_5 ] = Y_3$ . & {} \\ \end{tabular}
\medskip

Thus, the involution system $\mathcal{Y}$ is actually an algebra;
and the vector field we have to add are an ideal in $\mathcal{Y}$.

As for equations \eqref{eq:la} and \eqref{eq:lagen}, it is
immediate to check these are not satisfied: the r.h.s. of both
vanishes (in this case $\mu_{ij}^k = 0$ for all $i,j,k$), and the
l.h.s. of \eqref{eq:la} is easily checked to be nonzero. More
precisely, let us define $Q_{ij}^k = Y_i (\s_j^k) - Y_j (\s_i^k )$
(it is obvious that $Q_{ij}^k = - Q_{ji}^k$, and hence $Q_{ii}^k =
0$ for all $i$); then $Q_{12} = \( u_x , - v_x \)$. Applying this
on vectors $\phi_k$, i.e. computing the l.h.s. of
\eqref{eq:lagen}, we get again (the matrix built with the $\phi_i$
as columns is just the identity)
$$ Q_{ij}^k \phi^a_k \ = \ ( u_x , - v_x ) \ \not= \ 0 \ . $$

\subsubsection*{Case 2.}

Similar considerations apply if we consider the vector fields of
Example 2, i.e. $X_1 = u \pa_u$ and $X_2 = - u \pa_v$, and a
matrix which is the transpose of the one used there,
\beq\label{eq:matex3} \s \ = \ \pmatrix{0 & u_x \cr u & 0 \cr} \ .
\eeq In this case $ Y_1 = u \pa_u + u_x \pa_{u_x} - u u_x
\pa_{v_x}$; $Y_2 = u \pa_{v} + u^2 \pa_{u_x} - u_x \pa_{v_x}$. We
have then to introduce a vector field $Y_3 = u^3 \pa_{v_x}$ and
have
$$ [Y_1,Y_2] = Y_2 + Y_3 \ , \ \ [Y_1,Y_3] = 3 \, Y_3 \ , \ \  [Y_2,Y_3] = 0 \ . $$
In this case we get $Q_{12} = (u , - u^2)$; $Q_{12}^k  \phi^a_k =
(u^2 , u^3 )$.

\subsubsection*{Case 3.}

Finally, let us consider the vector fields of Example 1, $X_1 =
\pa_u$ and $X_2 = \pa_v$, and the matrix $\s$ just considered
above, see \eqref{eq:matex3}. In this case the first joint
prolongation is given by $Y_1 = \pa_{u} + u_x \pa_{v_x}$, $Y_2 =
\pa_{v} + u \pa_{u_x}$; in order to close the involution relations
we have to add two new vector fields in $J^1 M$, i.e. $Y_3 =
\pa_{u_x} - u \pa_{v_x}$; $Y_4 = \pa_{v_x}$. With these, the
involution (actually, again algebraic) relations are then

\smallskip
\begin{tabular}{lll}
$[Y_1,Y_2] = Y_3$ \ , & \ $[Y_1,Y_3] = - 2 \, Y_4$ \ , & \ $[Y_1,Y_4] = 0$ \ , \\
$[Y_2,Y_3] = 0$ \ , & \ $[Y_2,Y_4] = 0$ \ , & \ $[Y_3,Y_4] = 0$ \
. \end{tabular}
\medskip

As for equations \eqref{eq:la} and \eqref{eq:lagen}, again these
are not satisfied; the r.h.s. of both vanishes for $\mu_{ij}^k =
0$, and as for the l.h.s. we have $Q_{12} = \( 1 , - u \)$,
$Q_{ij}^k \phi^a_k = ( 1 , - u ) \ \not= \ (0,0)$.

\subsection*{Example 4.}

We aim now at illustrations of Theorem 5. Let us consider the
(obviously commuting) vector fields $X_1 = \pa_v$, $X_2 = (1/u)
\pa_u$; we $\s$-prolong them with the (nearly trivial) matrix $ \s
= (u_x /u ) \, I$ ($I$ being the two-dimensional identity matrix);
in this way we get $ Y_1 = \pa_v  +  (u_x / u) \pa_{v_x}$, $Y_2 =
 (1/u) \pa_u$.

We note that these are not in involution: in fact, $ [Y_1 , Y_2 ]
=  (u_x / u^3) \pa_{v_x} := Y_3$; this satisfies $ [Y_1,Y_3] = 0$,
$[Y_2 , Y_3] = - (3/u^2)  Y_3$, hence closes the algebra.

With the notation introduced in Example 3, we have $Q_{12} =
(u_x/u^3 , 0 )$; $Q_{12}^k \phi^a_k = (0,u_x/u^3)$.

We now look for a set of standardly prolonged vector fields which,
as stated by Theorem 5, are in the module generated by
$(Y_1,Y_2)$. As suggested by Theorem 5, we look for $A(x,u,v)$
solution to \eqref{eq:A}, i.e. in this case to
$$ D_x A \ = \ (u_x / u) \ A \ . $$
The solution to this equation is not unique: any matrix of
the form
$$ K \ = \ \pmatrix{k_1 \, u & k_2 \, u \cr k_3 \, u & k_4 \, u \cr} $$
with $k_i$ constants is a solution. Let us look in particular at
the matrices
$$ A \ = \ \pmatrix{0 & u \cr u & 0 \cr} \ , \ \
B \ = \ \pmatrix{u & 0 \cr 0 & u \cr} \ . $$

In the present case, using $A$ we get
$$ Z_1^{(A)} \ = \ u \, \pa_u \ + \ u_x \, \pa_{u_x} \ , \ \
Z_2^{(A)} \ = \ \pa_v \ ; $$
using $B$ we get instead
$$ Z_1^{(B)} \ = \ u \, \pa_v \ + \ u_x \, \pa_{v_x} \ , \ \
Z_2^{(B)} \ = \ \pa_u \ . $$
These sets are transformed one into the other by the matrix
$$ M \ = \ \pmatrix{0&1\cr 1&0\cr} \ , $$
which also maps $A$ and $B$ one into the other; more precisely,
writing
$$ Z_i^{(A)} = \eta^a_i \pa_a + \beta^a_i \pa_a^x \ , \ \
Z_i^{(B)} = \gamma^a_i \pa_a + \theta^a_i \pa_a^x \ , $$
we have
$$ \gamma_i^a = M^a_b \eta_i^b \ , \ \ \theta_i^a = M^a_b \beta_i^b \ . $$

We stress that
$$ [ Z_1^{(A)} , Z_2^{(A)} ] \ = \ 0 \ ; \ \
[ Z_1^{(B)} , Z_2^{(B)} ] \ = \ - \pa_v \ \not\in \
\mathrm{span} \( Z_1^{(B)} , Z_2^{(B)} \) \ . $$ On the other hand,
$Z_3^{(B)} = - \pa_v$ commutes with both $Z_1^{(B)}$ and $Z_2^{(B)}$.

Finally, let us look at invariants; a set of common invariants (beside $x$)
for $\{ Z_i^{(A)} \}$ is provided by $\{v_x , u_x/u \}$.
As for the vector fields $Z_i^{(B)}$, we need to add the
third vector field $Z_3^{(B)}$ and hence we expect (as is indeed
the case) that only one common differential invariant of order one
exists (beside $x$); this is $u_x$.

\subsection*{Example 5.}

We now illustrate Lemma 2, i.e. the converse of Theorem 5. Let us
consider the vector fields $W_1 = u \pa_u + v \pa_v $,  $W_2 = - v
\pa_u + u \pa_v$; their (standard) prolongations are of course
$$ Z_1 = u \pa_u + v \pa_v + u_x \pa_{u_x} + v_x \pa_{v_x} \, \ \
Z_2 = - v \pa_u + u \pa_v - v_x \pa_{u_x} + u_x \pa_{v_x} \ . $$

According to Lemma 2, if we operate on these with a linear
transformation $A(x,u,v)$ depending on variables in $M$, we should
obtain a set of vector fields which are a $\s$-prolonged set, with
$\s $ corresponding to $\s = A^{-1} (D_x A)$; more precisely, the
$\{Y_i \}$ should be the $\s$-prolongation of the $X_i = A_{ij}
W_j$.

We will write $\Phi = (1-uv)^{-1}$, and choose
$$ A \ = \ \pmatrix{u & 1 \cr 1 & v \cr} \ , \ \ A^{-1} \ = \
\Phi \ \pmatrix{- v & 1 \cr 1 & - u \cr} \ , $$
which yields
$$ \s \ = \ \Phi \ \pmatrix{- v u_x & v_x \cr
u_x & - u v_x \cr} \ . $$

With simple computations, we get
\begin{eqnarray*}
Y_1 &=& \Phi  \[ (1-u) v \pa_u  +  (1-v) u \pa_v  +
(v_x - v u_x) \pa_{u_x}  +  (u_x - u v_x) \pa_{v_x} \]  , \\
Y_2 &=& \Phi  \[ (u + v^2) \pa_u  -  (u^2 + v)
\pa_v  +  (u_x + v v_x) \pa_{u_x}  - (v_x + u u_x )
\pa_{v_x} \]  . \end{eqnarray*} These are indeed the
$\s$-prolongation, with the $\s$ given above, of the vector fields
\begin{eqnarray*}
 X_1 &=& A_{1j} \, W_j \ = \ \Phi \
 \[ (1-u) \, v \, \pa_u \ + \ (1-v) \, u \, \pa_v \] \ , \\
 X_1 &=& A_{1j} \, W_j \ = \ \Phi \
 \[ (u + v^2) \, \pa_u \ - \ (u^2 + v) \, \pa_v \] \ . \end{eqnarray*}

Let us now consider first order differential invariants; it is
easily checked that $Z_1$ admits as invariants $\{ x , u/v , u_x /
v_x \}$, while for $Z_2$ we get $\{ x , u^2+v^2 , u_x^2 +v_x^2
\}$; common differential invariants are
$$ P \ = \ \frac{u_x^2 + v_x^2}{u^2 + v^2} \ , \ \ Q \ = \
 \arctan \( u_x / v_x \) \, - \, \arctan \( u / v \)  \ . $$
Applying $Y_1$ and $Y_2$ on these, we obtain a non-zero result in
all cases, as stated in Remark \ref{sec:geo}.3.

\subsection*{Example 6.}

We give another illustration of Lemma 2, this time for vector
fields which are not defined in $u=v=0$. We write $\rho = (u^2 +
v^2)$ and choose
$$ W_1 \ = \ \rho^{-1} \, h (x,u) \, \pa_u \ , \ \
W_2 \ = \ \rho^{-1} \, h (x,v) \, \pa_v \ , $$ with $h$ an
arbitrary smooth function; this yields as first standard
prolongation the vector fields
\begin{eqnarray*} Z_1 &=& W_1 \ + \ \rho^{-2} \ \( \rho \, [D_x h(x,u)] \, - \,
h(x,u) \, (D_x \rho) \) \, \pa_{u_x} \ , \\
Z_2 &=& W_2 \ + \ \rho^{-2} \ \( \rho \, [D_x h(x,v)] \, - \, h(x,v) \,
(D_x \rho) \) \, \pa_{v_x} \ . \end{eqnarray*}

Let us now consider a transformation
$$ A \ = \ - \, \pmatrix{v & u \cr - u & v \cr} \ , \ \ A^{-1} \ = \
- \frac{1}{\rho} \ \pmatrix{v & u \cr u & v \cr} \ ; $$
and the vector fields
$$ X_i \ = \ A_i^{\ j} \ W_j \ , \ \ Y_i \ = \ A_i^{\ j} \ Z_j \ . $$
Writing $P = h(x,u)$, $Q = h (x,v)$; and moreover
$\mathcal{P} = \( \rho (D_x P)  - 2 (D_x \rho ) P \)$,
$\mathcal{Q} = \( \rho  (D_x Q) - 2 (D_x \rho ) Q \)$,
these are given explicitly by
\begin{eqnarray*}
X_1 &=& - \ \rho^{-1} [ v \, P \, \pa_u \ + \
u \, Q \, \pa_v ] \ , \\
X_2 &=&  \rho^{-1} \ [ u \, P \, \pa_u \ - \
v \, Q \, \pa_v ] \ ; \\
Y_1 &=& X_1 \ - \  \rho^{-2} \[ v \mathcal{P} \, \pa_{u_x} \ - \ u \mathcal{Q}  \,
\pa_{v_x} \] \ , \\
Y_2 &=& X_2 \ + \ \rho^{-2} \[ u \mathcal{P} \, \pa_{u_x} \ - \ v \mathcal{Q}  \,
 \pa_{v_x} \] \ . \end{eqnarray*}

It is easily checked that the $\{ Y_i \}$ are the
$\s$-prolongation of the $\{ X_i \}$ with
$$ \s \ = \ \frac{1}{\rho} \ \pmatrix{- (u u_x + v v_x) & u v_x - v u_x \cr
- u v_x + v u_x & - (u u_x + v v_x) \cr} \ = \ A \ (D_x A^{-1} ) \ . $$

\subsection*{Example 7.}

In the previous example, both the $Y_i$ and the $Z_i$ fields were
singular. Situations where the standardly prolonged fields are
singular but the $\s$-prolonged ones are regular would be of
interest; here we deal with such a situation, obtained through a
small variation on the setting of Example 6. We use the same
notation introduced there.

We consider the (singular) vector fields
$$ X_1 \ = \ \( P / \rho\) \, \pa_u \ , \ \
X_2 \ = \ \( Q / \rho \) \, \pa_v \ ; $$ these have standard
prolongations \begin{eqnarray*} Z_1 &=& X_1 \ + \ \[ ( \rho
(D_x P) - P (D_x \rho) ) / \rho^2 \] \, \pa_{u_x} \ , \\
Z_2 &=& X_2 \ + \ \[ ( \rho (D_x Q) - Q (D_x \rho) ) / \rho^2 \]
\, \pa_{v_x} \ . \end{eqnarray*} Acting now with
$$ A \ = \ \rho^2 \ \pmatrix{v & - u \cr u & v \cr } $$
we get the (regular) vector fields
\begin{eqnarray*} X_1 &=& \rho  \[ ( P v) \pa_u -
( Q u) \pa_v \] \ , \ \ X_2 \ = \ \rho \[ ( P  u) \pa_u  +
( Q v) \pa_v \] \ ; \\
Y_1 &=& X_1 +   v [\rho (D_x P) - P (D_x \rho) ] \pa_{u_x} -
u[\rho (D_x Q) - Q (D_x \rho) ] \pa_{v_x}  \ , \\
Y_2 &=& X_2 + u [\rho (D_x P) - P (D_x \rho) ] \pa_{u_x} + v [\rho
(D_x Q) - Q (D_x \rho) ] \pa_{v_x} \ .
\end{eqnarray*}

One can check that the $\{ Y_i \} $ are the $\s$-prolongation of
the $\{ X_i \}$, with
$$ \s \ = \ A \, (D_x A^{-1} ) \ = \ - (D_x A) \, A^{-1} \ = \
- \, \frac{1}{\rho} \ \pmatrix{5 (u u_x + v v_x) & u v_x - v u_x
\cr - u v_x + v u_x & 5 (u u_x + v v_x ) \cr} \ \ \ . $$

\subsection*{Example 8.}

Let us now see an example of the situation considered in Theorem
4; that is, we will have a set of three differential equations of
second order, admitting as symmetries a set of two $\s$-prolonged
vector fields in involution (actually, commuting). (A
number of trivial examples are also easily obtained by adding to
any $n$-dimensional example for theorem 3 (with $\Y$ of rank $n$)
a new equation for a new dependent variable $w (x)$; this is not
acted upon by, nor entering in, the coefficients of the considered
vector fields and hence no reduction is possible on it. We will
thus end up with a system of one second order and $n$ first order
equations.)

We consider the equations
\begin{eqnarray*} E_1 &:=&  e^{-(u
+ w)} \left( u_{xx} - v_{xx} \right) \,
     \left( u_{x} - v_{x} - w_{x} \right) \, - \,
     [{u_{x} + w_{x}}]^{-1} \, (u - v ) \, \times \\
     & & \ \times \, ( -{u_{x}}^2 + u_{xx} + u_{x}\,v_{x} +
       v_{x}\,w_{x} + {w_{x}}^2 + w_{xx} )
        \ = \ 0 \
; \\ E_2 &:=& e^{-(u - v - w)} \, ( u_x - v_x ) \,
   \left( u_{xx} + w_{xx} - u_x^2  + u_x v_x + v_x w_x +
     w_x^2  \right)  + \\
     & & \ - \ [{u_{x} - v_{x} -
     w_{x}}]^{-1} \, ( u - v ) \, \times \\
     & & \ \times \left( -{u_{x}}^2 + u_{xx} +
u_{x}\,v_{x} - v_{xx} +
       v_{x}\,w_{x} + {w_{x}}^2 - w_{xx} \right)   \ = \ 0 \ ; \\
E_3 &:=& (u-v) \ - \ e^{-(u+w)} \,\left( u_x - v_x \right) \
\times \\
 & & \ \times \
   \left( u_{xx} - v_{xx} - w_{xx} - u_x^2 + u_x v_x  +
     v_x w_x + w_x^2  \right)
 \ = \ 0 \ ; \end{eqnarray*}
  and the (autonomous) commuting vector fields
$$ X_1 = \pa_u + \pa_v - \pa_w \ , \ \ X_2 = \pa_u + \pa_v  \ . $$
As for their (second) $\s$-prolongation, we set
$$ \s \ = \ \pmatrix{0 & u_x + w_x & 0 \cr u_x - v_x - w_x & 0 & 0 \cr 0&0&0\cr} $$
and hence we get, with standard computations, \begin{eqnarray*}
Y_1 &=& (\pa_u + \pa_v - \pa_w ) \ + \  (u_x + w_x) \, (\pa_{u_x}
+ \pa_{v_x} ) \\ & & \ + \ [u_{xx} + w_{xx} + (u_x - v_x - w_x)
(u_x + w_x)] \, (\pa_{u_{xx}} + \pa_{v_{xx}} ) \\ & & \ - \ (u_x -
v_x -
w_x) (u_x + w_x) \, \pa_{w_{xx}} \ ; \\
Y_2 &=& (\pa_u + \pa_v) \ + \ (u_x - v_x - w_x ) \, (\pa_{u_x} +
\pa_{v_x} - \pa_{w_x} \\ & & \ + \ [u_{xx} - v_{xx} + (u_x - v_x -
w_x) (u_x + w_x) - w_xx] \, (\pa_{u_{xx}} + \pa_{v_{xx}}) \\
& & \ - \ (u_{xx} - v_{xx} - w_{xx} ) \, \pa_{w_{xx}} \ .
\end{eqnarray*}

Now, by trivial dimension counting, there must be two invariants
of order zero, i.e. on $M$; these are
$$ \eta_1 \ = \ x \ , \ \ \eta_2 \ = \ u \, - \, v \ . $$
The common differential invariants of order one for these vector
turn out to be
$$ \z_1^{(0)} = e^{- (u - v - w)} \, (u_x + v_x) \ , \ \
\z_2^{(0)} = e^{- (u+w)} \, (u_x - v_x - w_x )  \ , \ \ \z_3^{(0)}
= u_x - v_x \ . $$ The second order differential invariants can be
readily computed by the IBDP (using $\eta_1$) as $\zeta_i^{(1)} =
D_x \zeta_i^{(0)}$, which yields \begin{eqnarray*}
 \z_1^{(1)} &=& e^{- (u-v-w)} \, \( u_{xx} + w_{xx} + u_x v_x + v_x w_x - u_x^2 +
w_x^2 \) \ , \\
\z_2^{(1)} &=& e^{- (u+v)} \, \( u_{xx} - v_{xx} - w_{xx} + u_x
v_x + v_x w_x - u_x^2 + w_x^2 \) \ , \\
\z_3^{(1)} &=& u_{xx} - v_{xx} \ . \end{eqnarray*} these are
easily checked to be indeed invariant under the $Y_i$.

We can then perform the change of variables
$$ \xi = u - v \ , \ z_1 = \zeta_1^{(0)} \ , \ z_2 = \zeta_2^{(0)}
\ ; $$ this of course entails a corresponding change for
derivatives, which we do not write down explicitly.
In the new variables, the equations are written as
\begin{eqnarray*} E_1 &:=& \xi \, \( (z_1)_x / z_1 \)
\ - \ z_2 \, \xi_{xx} \ = \ 0 \ ; \\
E_2 &:=& \xi \, \( (z_2)_x / z_2 \) -  \xi_x \, (z_1)_x \ = \ 0 \
; \\
E_3 &:=& \( (z_2)_x / z_2 \) \ - \ \( (z_1)_x / z_1 \) \ = \ 0 \ .
\end{eqnarray*} The system is now of first order in the $z_i$
variables, and of second order in $\xi$; it can be rewritten as
$$
\frac{d^2 \xi}{d x^2} \, = \, \frac{\xi^3}{\xi_x^2 \, z_1 \, z_2^2}
\ , \ \ \frac{d z_1}{d x} \, = \, \frac{\xi^2}{\xi_x^2 \, z_2} \ ,
\ \ \frac{d z_2}{d x} \, = \, \frac{\xi}{\xi_x} \ . $$

Needless to say, the same procedure -- at least up to rewriting
the system as first order in the $z_i$ variables, and second order
in $\xi$ --  would work for any starting system of equations in
the form $ E_i := F_i \[ \eta_1,\eta_2;
\z_1^{(0)},\z_2^{(0)},\z_3^{(0)}; \z_1^{(1)},\z_2^{(1)},\z_3^{(1)}
\] = 0$, leading to equations written in the new
variables as $ F_i [ x, \xi; z_1, z_2, \xi_x; (z_1)_x, (z_2)_x,
\xi_{xx} ] = 0$.

\subsection*{Example 9.}

Let us now consider, again in order to illustrate Theorem 4, a
case with three dependent variables $(u,v,w)$ and two (non
commuting) independent vector fields in involution,
$$ X_1 \ = \ u \, \pa_u \ + \ w \, \pa_w \ ; \ \
X_2 \ = \ - \, u \, \pa_v \ ; \ \ [X_1 , X_2 ] = X_2 \ . $$

We will use the $(2 \times 2)$ matrix
$$ \s \ = \ \pmatrix{0 & u_x \cr 0 & 0 \cr} \ . $$
With this choice, the $\s$-prolongations of the vector fields $\{
X_1 , X_2 \}$ to $J^2 M$ are
\begin{eqnarray*}
Y_1 &=& u \, \pa_u \ + \ w \, \pa_w \ + \ u_x \, \pa_{u_x} \ - \ u \, u_x \, \pa_{v_x} \ + \ w_x \, \pa_{w_x} \\
 & & \ + \ u_{xx} \, \pa_{u_{xx}} \ - \ (2 \, u_x^2 \, + \, u \, u_{xx}) \, \pa_{v_{xx}} \ + \ w_{xx} \, \pa_{w_{xx}} \ , \\
Y_2 &=& - \, u \, \pa_v \ - \ u_x  \, \pa_{v_x} \ - \ u_{xx} \,
\pa_{v_{xx}} \ . \end{eqnarray*} These still satisfy the same
commutation (hence involution) relation: $ [ Y_1 ,Y_2 ] = Y_2$.

With trivial algebra one finds the common invariants of order up
to two for these two vector fields; by trivial dimension counting
they must be eight, one of them being the trivial one $J_{00} =x$.

A simple basis for the (seven) nontrivial ones is given by:
$J_{01} = (w/u)$, $J_{11} = (w_x/u)$, $J_{21} = (w_{xx}/u)$;
$J_{12} = (u_x / u)$, $J_{22} = (u_{xx} / u)$; $J_{13} = [v_x  + (
u  u_x / 2)  - (u_x  v / u)]$, $J_{23} = [v_{xx} +  u_x^2  +  (u
u_{xx}/2)  - (u_{xx}  v / u)]$. Another possible, maybe slightly
less simple but more convenient, basis is provided by
\begin{eqnarray*}
\zeta_1^{(0)} &=& w/u \ ; \ \ \zeta_1^{(1)} \ = \  (u w_x \, - \, u_x w)/u^2 \ , \ \ \zeta_2^{(1)} \ = \ u_x/u \ , \\
\zeta_3^{(1)} &=& v_x \, + \, u u_x / 2 - u_x v/ u \ ; \\
\zeta_1^{(2)} &=& (u^2  w_{xx} \, - \, u  w  u_{xx} \, - \, 2 u u_x w_x \, + \, 2 w u_x^2)/u^3 \ , \\
\zeta_2^{(2)} &=& (u u_{xx} - u_x^2)/u^2 \ , \ \ \zeta_3^{(2)} \ =
\ (v_x \, + \, u u_x / 2 - v u_x/ u) \ . \end{eqnarray*} This is
convenient in that (wherever applicable) \beq\label{eq:9.1}
\zeta_a^{(i+1)} \ = \ D_x \zeta_a^{(i)} \ . \eeq

Thus any system of equations $E^{(i)}: F^{(i)}
[x;u^{(2)},v^{(2)},w^{(2)}] = 0$ of order not higher than two
written in terms of these joint invariants will admit the $\{ X_1
, X_2 \}$ as $\s$-symmetries (with the $\s$ given above), and
conversely. Note all of these (if nontrivial) will necessarily be
singular for $u=0$.

In order to consider a concrete example, let us look at the system
\begin{eqnarray*}
u_{xx} &=& \frac{2 w u_x^2 + u^2 (2 v u_x - 2 u v_x + w_x - w) -
u u_x (u^3+w+w_x) }{u  w} , \\
v_{xx} &=& \frac{ A  + B   +  C }{2 \, u^2 \, w} \ , \\
w_{xx} &=& - \ \frac{(u + u_x) \, (w - w_x)}{u} \ ; \\
A &=& 2 u^5 v_x - ((4 v + u^2) u_x - w + w_x ) u^4
+ (u_x (w+w_x)- 4 v v_x) u^3  , \\
B &=& [(-3 w  u_x^2 + 4 v^2 u_x +2 v (w_x-w)] u^2 \ , \\
C &=& 2 (w w_x - u_x (v (w+w_x)-v_x w))
u + 2  u_x^2 v w \ .
\end{eqnarray*}

In terms of the invariants, this system reads simply
$$ \begin{array}{l}
\zeta_2^{(2)} - \zeta_1^{(1)} + \zeta_1^{(0)} \ = \ 0 \
, \\
\zeta_3^{(2)} \ - \ \zeta_2^{(1)}  \ = \ 0 \ , \\
\zeta_1^{(2)} \ - \ 2 \, \zeta_3^{(1)} \ = \ 0 \ . \end{array} $$

In order to reduce it to a system of one second order equation and
two first order ones, we should change (dependent) coordinates. We
choose \beq\label{eq:9.2} \xi \ = \ \zeta_1^{(0)} \ , \ \ \eta \ =
\ \zeta_2^{(1)} \ , \ \ \rho \ = \ \zeta_3^{(1)} \ ; \eeq and
recall \eqref{eq:9.1}. The equations are then readily expressed in
the new coordinates, providing \beq \frac{d^2 \xi}{d x^2} \ = \ 2
\, \rho \ , \ \ \frac{d \rho}{d x} \ = \ \eta \ , \ \ \frac{d
\eta}{d x} \ = \ \frac{d \xi}{d x} \ - \ \xi \ ; \eeq i.e., as
claimed, a system of two first order and one second order
equations.

\vfill\eject

\section*{Appendix A.
Relations between $\s$- and $\mu$-prolongations}

The picture emerging from the discussion of Section \ref{sec:geo}
above, see in particular Theorem 5 and Lemma 2, is quite
reminiscent of the one holding for $\mu$-prolongations
\cite{CGM,Gtwist} (here we assume the reader to be familiar with
them; see \cite{Gtwist} for details and a bibliography on twisted
prolongations in general): in this appendix we will discuss
similarities and differences between the two, and their relations
when applied to a given set of vector fields.

\subsection*{A1. Similarities and differences}

First of all, let us stress the analogies between $\s$- and
$\mu$-prolongations: in both cases,
 \begin{itemize}
 \item[(a)] the
vector fields $Y_i$ obtained from $X_i$ through the modified ($\s$
or $\mu$) prolongation operation are the standard prolongations
$Z_i$ of different vector fields $W_i$;
 \item[(b)] the relation between $X_i$
and $Y_i$ on the one hand, and $W_i$ and $Z_i$ on the other, can
be summarized by the action of a linear transformation $A$;
 \item[(c)] this $A$ then determines the relevant matrix ($\s$ in our case, a
single $\La$ in the case of $\mu$-prolongations for ODEs) entering
in the definition of the modified prolongation operation -- and
reducing to zero for standard prolongations.
\end{itemize}

An essential difference between the two cases should however be
emphasized: in the case of $\mu$-prolongations, the linear
transformation acts on the basis vectors in $\T M$ and more
generally in $\T (J^k M)$ \cite{Gtwist}; here instead we are
changing the generators of the module of vector fields -- i.e. act
on a structure superimposed to $\T (J^k M)$ with no action on the
space itself.

In other words, in the case of $\mu$-prolongations {\it all }
vector fields on $M$ and on $J^k M$ are affected by the action of
the linear transformation -- as obvious since this is a
point-dependent change of basis, i.e. a gauge transformation
\cite{Gtwist} -- while here we are just changing our choice for
the generators of a given module of vector fields, or equivalently
the generators for a given distribution on $\T M$ and $\T (J^k
M)$, with of course no action on the general set of vector fields
in $M$ or $J^k M$.

As a consequence, the $\mu$-prolongation operation acts
individually on each vector field; they are thus well defined, and
indeed different from standard ones, also for single vector
fields. Moreover -- in the framework of ODEs, which is the only
one of interest here -- $\mu$-prolongations reduce to
$\la$-prolongations only if the linear transformation $A$ is a
multiple of the identity, the proportionality factor being a
smooth function on $M$.

On the other hand, the $\s$-prolongation operation is defined on
{\it finite sets} of vector fields, and reduces to a somewhat
trivial one in the case where the set mentioned above is made of a
single vector field. They again reduce to usual
$\la$-prolongations only if the linear transformation $A$ is a
multiple of the identity, the proportionality factor being a
smooth function on $M$.

We summarize the situation in the following commutative diagram:
\beq\label{diag:svsmu} \matrix{ \{ \^X_i \} & \mapleft{M} & \{ W_i
\} & \mapright{S} & \{ X_i \} \cr
 & & & & \cr
 \mapdown{\mu-\mathtt{prol}} & & \mapdown{\mathtt{prol}} & &
 \mapdown{\s-\mathtt{prol}} \cr
 & & & & \cr
\{ \^Y_i \} & \mapleft{M} & \{ Z_i \} & \mapright{S} & \{ Y_i \}
\cr} \eeq Here $\^Y_i$ is the $\mu$-prolongation of $\^X_i$, the
$\{ Y_i \}$ are the $\s$-prolongation of the $\{ X_ i \}$, and the
inertible linear transformations $S$ and $M$ are related to $\s$
and to $\mu = \La \d x$ via \beq \La \ = \ M \, (D_x M^{-1}) \ , \
\ \s \ = \ S \, (D_x S^{-1}) \ . \eeq

\subsection*{A2. Relations}

It is quite clear that albeit the two operations are conceptually
different in general, a relation exists between the two points of
view when we consider a given transformation on a given set of
vector fields; this is quite similar to the relation between
``active'' and ``passive'' points of view in fluid mechanics.

We discuss it for vertical vector fields only (they can be thought
as evolutionary representatives of general vector fields in $M$,
see \cite{\symmref}), and refer to \eqref{diag:svsmu} for
notation.

Consider vector fields $W_i$ in $M$ and their prolongations $Z_i$
in $J M$; consider also a linear (point-dependent) transformation
in $M$, under which $W_i$ are mapped into $X_i$ and $Z_i$ into
$Y_i$; we know that the $\{ Y_i \}$ are then the $\s$-prolongation
of the $\{X_i \}$. We want to discuss if there is a linear map $M$
such that these particular $W_i$ and $Z_i$ are mapped into $\^X_i$
and $\^Y_i$ (each of them being the $\mu$-prolongation of the
corresponding $\^X_i$), with the additional properties that {\tt
(a)} $\^X_i = X_i$; {\tt (b)} $\^Y_i = Y_i$. It will suffice to
work on first prolongations, as prolongations of order $k+1$ can
be seen as first prolongations of prolongations of order $k$.

It is convenient to work in coordinates, writing $W_i = \phi^a_i
\pa_a$, and embodying the components of the different vector
fields into a single matrix $\Phi$ with elements
\beq\label{eq:phimat} \Phi^a_{\ i} \ = \ \phi^a_i \ . \eeq Note
that if we have $r$ vector fields with $n$ components, the matrix
$\Phi$ is $(n \times r)$. In particular, it is a square -- and
thus possibly invertible -- matrix if and only if $r=n$; this is
precisely the case we have considered in the main body of the
present paper. Note also that if $r=n$, the condition for $\Phi$
to be invertible (at all points) is precisely that the vector
fields $X_i$ are independent (at all points).

Coming back to our vector fields, $X_i$ has components
$\wt{\phi}^a_i = (S_i^{\ j} \phi^a_j$, while $\^X_i$ has
components $\^\phi^a_i = M^a_{\ b} \phi^b_i$. Requiring the two to
be the same, i.e. $\wt{\phi} = \^\phi$, amounts to requiring
$$ S_i^{\ j} \, \phi^a_j \ = \ M^a_{\ b} \, \phi^b_i \ ; $$ or,
acting with $M^{-1}$ from the left,
$$ \phi^b_i \ = \ (M^{-1})^b_{\ a} \, S_i^{\ j} \, \phi^a_j \ = \
(M^{-1})^b_{\ a} \, \phi^a_j \, (S^T)^j_{\ i} \ . $$ Rearranging
the summation indices and using the notation \eqref{eq:phimat},
this is also rewritten as the matrix equation
\beq\label{eq:srelmu} \Phi \ = \ M^{-1} \, \Phi \, S^T \ . \eeq

It should be recalled that in our discussion of $\s$-prolongations
we have supposed the vector fields $X_i$ to be as many as the
dependent variables, and to be independent {\it at all points} of
$M$; in view of the requirement that also $S$ and $M$ are
invertible at all points, this means that the $W_i$ will also be
independent, and hence $\Phi$ will be nonsingular, at all points
of $M$.

Thus if we are requiring the $M$ which satisfies \eqref{eq:srelmu}
for given $\Phi$ and $S$, the answer is that this is given by
\beq\label{eq:musrelmu} M \ = \ \Phi \, S^T \, \Phi^{-1} \ ; \eeq
similarly, if $\Phi$ and $M$ are given and we ask if there is a
$S$ such that \eqref{eq:srelmu} is satisfied, we get \beq S \ = \
\Phi^T \, M^T \, (\Phi^{-1})^T \ , \ \ S^T \ = \ \Phi^{-1} \, M \,
\Phi \ . \eeq Finally, if $M$ and $S$ are given and we wonder
which are the vector fields $W_i$ such that $X_i = \^X_i$, the
answer is provided directly by \eqref{eq:srelmu}.

Our discussion so far only concerns vector fields in $M$; let us
now consider also vector fields in $J M$. We write $Z_i$ as
$$ Z_i \ = \ \phi^a_i \, (\pa / \pa u^a) \ + \ \psi^a_i \, (\pa /
\pa u^a_x ) \ ; $$ here of course $\psi^a_i = D_x \phi^a_i$. We
will embody the coefficients $\psi^a_i$ into a matrix $\Psi$ with
entries $\Psi^a_{\ i} = \psi^a_i$; this satisfies $\Psi = D_x
\Phi$.

Proceeding in exactly the same way as above, we easily get that
$$ Y_i \ = \ X_i \ + \ [ S_i^{\ j} (D_x \phi^a_j)] \, (\pa / \pa
u^a_x) \ , \ \ \^Y_i \ = \ \^X_i \ + \ [ M^a_{\ b} (D_x \phi^b_i)]
\, (\pa / \pa u^a_x) \ ; $$ thus -- assuming $X_i = \^X_i$ -- we
have to require
$$ \Psi \, S^T \ = \ M \, \Psi \ . $$
Recalling now that $\Psi = D_x \Phi$, this yields with trivial
manipulations $$ (D_x M ) \ = \ \Phi \, (D_x S^T ) \, \Phi^{-1}
\ ; $$ recalling also \eqref{eq:musrelmu}, we arrive at the
condition \beq\label{eq:svsmuprol} [\Phi^{-1} \, (D_x \Phi)] \ S^T
\ = \ S^T \ [\Phi^{-1} \, (D_x \Phi)] \ . \eeq

In other words, the equality $X_i = \^X_i$ will be lifted to a
corresponding equality $Y_i = \^Y_i$ between prolonged vector
fields if and only if \eqref{eq:svsmuprol} is satisfied.

\subsection*{A3. Combining $\mu$- and $\s$-prolongations}

The previous discussion also suggests how to combine $\mu$- and
$\s$-prolongations into a single prolongation operation; we will
call it {\it $\chi$-prolongation} (where $\chi$ stands for
``combined''): working with matrices $\Phi$ and $\Psi = D_x \Phi$,
and writing $R = S^T$ for ease of notation, one should pass to
\beq\label{eq:mixprol1} \^\Phi \ = \ M^{-1} \, \Phi \, R \ , \ \
\^\Psi \ = \ M^{-1} \, \Psi \, R \ ; \eeq the operator mapping
$\^\Phi$ into $\^\Psi$ will be that of the prolongation combining
$\mu$ and $\s$ ones. Inverting the first of the relations
\eqref{eq:mixprol1}, we have
$$ \Phi \ = \ M \, \^\Phi \, R^{-1} \ , $$ hence the second of
\eqref{eq:mixprol1} reads \begin{eqnarray*}
 \^\Psi &=& M^{-1} [ D_x (M \, \^\Phi \, R^{-1})] \, R \\
  &=& M^{-1} \, [ (D_x M) \^\Phi R^{-1} \, + \, M (D_x \^\Phi) R^{-1} \,
  + \, M \^\Phi (D_x M^{-1} ] \, R \\
  &=& (D_x \^\Phi) \ + \ [M^{-1} \, (D_x M)] \, \^\Phi \ - \
  \^\Phi \, [R^{-1} \, (D_x R)] \\
  &=& (D_x \^\Phi) \ + \ \La \, \^\Phi \ - \
  \^\Phi \, \Theta \ ; \end{eqnarray*}
here we have of course defined
$$ \La \ := \ M^{-1} \, (D_x M) \ , \ \ \Theta \ := \ R^{-1} \, (D_x
R)\ . $$ Reintroducing indices and passing to the vector notation,
we rewrite the above relation as \beq \^\psi^a_i \ = \ (D_x
\phi^a_i) \ + \ \La^a_{\ b} \, \phi^b_i \ - \ (\Theta^T)_i^{\ j}
\, \phi^a_j \ ; \eeq this defines the (first, and then recursively
higher order) prolongation operator.

This can be described in terms of commutative diagrams by {\it
complementing} (not substituting!) the diagram \eqref{diag:svsmu}
with the diagram \beq\label{diag:chiprol} \matrix{ \{ \^X_i \} &
\mapright{S} & \{ P_i \} & \mapleft{M} & \{ X_i \} \cr
 & & & & \cr
 \mapdown{\mu-\mathtt{prol}} & & \mapdown{\chi-\mathtt{prol}} & &
 \mapdown{\s-\mathtt{prol}} \cr
 & & & & \cr
\{ \^Y_i \} & \mapright{S}  & \{ Q_i \} & \mapleft{M} & \{ Y_i \}
\cr} \eeq

One could thus think of investigating $\chi$-symmetries of
differential equations, defined as sets of vector fields which,
when $\chi$-prolonged, give sets of vector fields which leave the
solution manifold of a given set of ordinary differential
equations invariant.

However, it is well known that while $\mu$-symmetries are ``as
useful as standard symmetries'' in the search for special
solutions to PDEs or systems of PDEs, and they reduce to usual
$\la$-symmetries for scalar ODEs -- and are thus in a way again as
useful as standard ones in their reduction -- they are of no
(known) use in the case of systems of ODEs. Thus at the moment we
cannot describe any use of such $\chi$-prolongations, and they
remain at this stage a mathematical curiosity.

\vfill\eject

\section*{Appendix B. Determining equations for $\s$-symmetries}

In the Conclusions (Sect. \ref{sec:discussion} above) we have mentioned that the task of determining all $\s$-symmetries for a given equation or system is in general well beyond reach. In this Appendix we will illustrate this statement by a concrete example, i.e. the system considered in Example 1 above. This reads
\begin{eqnarray} u_{xx} &=& u_x \, v_x \ (1 + e^{-u}) \ , \nonumber \\
v_{xx} &=& u_x \, v_x \ (1 + e^{-v}) \ . \label{eq:B0} \end{eqnarray}
We will look for a set of {\it two} vector fields in involution $X_\i$ giving a $\s$-symmetry, and in order to keep formulas not too long, we will search for these with $\xi_\i = 0$. In this case the second prolongations will be written as
$$ Y_\i \ = \ \phi^a_\i \pa_a \ + \ \psi^a_\i \pa_a^{(x)} \ + \ \chi^a_\i \pa_a^{(xx)} \ , $$ where we have denoted $\pa_a^{(x)} := (\pa / \pa u^a_x )$ and $\pa_a^{(xx)} := (\pa / \pa u^a_{xx} )$; and the coefficients entering in first and second prolongations are now denoted as $\psi$ and $\chi$ to reduce the number of indexes.
An explicit computation shows that
\begin{eqnarray*}
\psi^a_\i &=& D_x \phi^a_\i \ + \ \s_{ij} \, \phi^a_\j \ , \\
\chi^a_\i &=& D_x \psi^a_\i \ + \ \s_{ij} \, \psi^a_\j  \\
 &=& D_x^2  \phi^a_\i \ + \ 2 \, \s_{ij} \, (D_x \phi^a_\j ) \ + \ [ D_x \s_{ij} \, + \, (\s^2)_{ij} ] \, \phi^a_\j \ . \end{eqnarray*}

In order to obtain the determining equations for $\s$-symmetries we should apply both prolonged vector fields to both equations, obtaining in all cases zero upon restriction to the solution manifold of the system \eqref{eq:B0}.

The determining equations -- under the present simplifying assumption $\xi_\i = 0$ -- are thus in this case (here and below $i=1,2$, $a=1,2$, and $S$ represents the solution manifold to the system)
\beq\label{eq:B1}
\left[ \chi^a_\i \ - \ [ \psi^1_\i \, v_x \ + \ u_x \, \psi^2_\i ] \, (1 + e^{-u^a}) \ + \ (u_x \, v_x) \, \phi^a_\i \, e^{- u^a} \right]_S \ = \ 0 \ . \eeq

Substituting according to the above for $\psi$ and $\chi$, and making explicit the derivative terms, we get (with $\phi^a_{\i,b} = \pa \phi^a_\i / \pa u^b$, etc.; sum over repeated indices is implied) for the left hand side of \eqref{eq:B1} before restriction to $S$
\begin{eqnarray*}
\( \phi^a_{\i,xx} \ + \ 2 \, u^b_x \, \phi^a_{\i,bx} \ + \ u^b_{xx} \, \phi^a_{\i,b} \ + \ u^b_x \, u^c_x \, \phi^a_{\i,bc} \) & & \\
\ + \ 2 \, \s_{ij} \, \( \phi^a_{\i,x} \, + \, u^b_x \, \phi^a_{\i,b} \) & & \\
\ + \ \[ \frac{\pa \s_{ij}}{\pa x} \, + \, u^a_x \, \frac{\pa \s_{ij}}{\pa u^a} \, + \, u^a_{xx} \, \frac{\pa \s_{ij}}{\pa u^a_x} \ + \ \s_{im} \s_{mj} \] \, \phi^a_\j \ . & & \end{eqnarray*}
Using now \eqref{eq:B0} to restrict on $S$, we obtain
\begin{eqnarray}
\( \phi^a_{\i,xx} \ + \ 2 \, u^b_x \, \phi^a_{\i,bx} \ + \ (u^1_x u^2_x (1 + e^{- u^b} )) \, \phi^a_{\i,b} \ + \ u^b_x \, u^c_x \, \phi^a_{\i,bc} \) & & \nonumber \\
\ + \ 2 \, \s_{ij} \, \( \phi^a_{\i,x} \, + \, u^b_x \, \phi^a_{\i,b} \) & & \label{eq:B2} \\
+ \ \[ \frac{\pa \s_{ij}}{\pa x} \, + \, u^a_x \, \frac{\pa \s_{ij}}{\pa u^a} \, + \, (u^1_x u^2_x (1 + e^{- u^a} )) \, \frac{\pa \s_{ij}}{\pa u^a_x} \ + \ \s_{im} \s_{mj} \] \, \phi^a_\j &=& 0 \ . \nonumber \end{eqnarray}

These are the determining equations we were looking for. It should be noted that they cannot be solved according to the standard procedure for determining equations of Lie-point symmetries, as they depend on the unknown matrix function $\s = \s (x,u,u_x)$.

The best one can do is to look for special solutions of these; e.g., if we look for solutions such that $\phi^a_\i$ are constant, and $\s$ only depends on $u_x$ and $v_x$, this reduces to
\beq
\[ (u^1_x u^2_x (1 + e^{- u^a} )) \, \frac{\pa \s_{ij}}{\pa u^a_x} \ + \ \s_{im} \s_{mj} \] \, \phi^a_\j \ = \ 0 \ . \label{eq:B4} \eeq
Despite its innocent-looking shape (due to compact notation) this is still a system of four coupled nonlinear PDEs.

We will now look for $\s$ in the form
$$ \s \ = \ \pmatrix{0 & A (u_x,v_x) \cr B(u_x , v_x ) & 0 \cr} \ ; $$ we also write $ \phi^1_{(1)} = c_1$, $\phi^2_{(1)} = c_2$, $\phi^1_{(2)} = k_1$, $\phi^2_{(2)} = k_2$;
in this way the determining equations \eqref{eq:B4} read
\begin{eqnarray*}
A e^v (-B c_1 e^u + (1 + e^u) (k_2 u_x + k_1 v_x)) & & \\
- (A_{v_x} e^u (1 + e^v) k_1 + e^v (c_1 + A_{u_x} (1 + e^u) k_1)) u_x v_x &=& 0 , \\
A e^u (-B c_2 e^v + (1 + e^v) (k_2 u_x + k_1 v_x)) & & \\
- (c_2 e^u + A_{u_x} (1 + e^u) e^v k_2 + A_{v_x} e^u (1 + e^v) k_2) u_x v_x &=& 0 ; \\
B e^v (-A e^u k_1 + (1 + e^u) (c_2 u_x + c_1 v_x))  & & \\
- (B_{v_x} c_1 e^u (1 + e^v) + e^v (B_{u_x} c_1 (1 + e^u) + k_1)) u_x v_x &=& 0 , \\
B e^u (-A e^v k_2 + (1 + e^v) (c_2 u_x + c_1 v_x))  & & \\
- (B_{u_x} c_2 (1 + e^u) e^v + B_{v_x} c_2 e^u (1 + e^v) + e^u k_2) u_x v_x &=& 0 .  \end{eqnarray*}
Determining the general solutions of these is not easy despite the several ansatzes considered to simplify them from the original form. A further reduction is obtained e.g. looking for $A,B$ as linear functions, $A(u_x,v_x) = p_1 u_x + p_2 v_x$, $B(u_x,v_x) = q_1 u_x + q_2 v_x$. In any case, it can be checked that the vector fields $X_1,X_2$ and the matrix $\s$ considered in Example 1 satisfies these equations.

Note that even assuming $\s = \s (x,u)$ (i.e. that $\s_{ij}$ are functions on $M$ and not on the full $J^1 M$), albeit in principles one gets a more tractable problem, in that all dependencies on derivatives are explicit and the determining equations can be decomposed into equation for the vanishing of different monomials in the $u^a_x$, the resulting PDEs are nonlinear in the $\s_{ij}$, and hence in general cannot be fully solved.
E.g., in the case of equations \eqref{eq:B0} even with the simplifying assumptions $\phi_\i = \phi_\i (u,v)$ and $\s_{ij} = \s_{ij} (u,v)$, after easily determining that $X_\i$ must be of the form
$$ X_\i = \[ \frac{\exp [(1 + e^{-u}) v]  }{1 + e^{-u}} \, P_\i (u) \ + \ Q_\i (u) \] \pa_u \ + \ \[ \frac{\exp [(1 + e^{-v}) u]  }{1 + e^{-v}} \, R_\i (u) \ + \ S_i (u) \] \pa_v \ , $$ one remains with PDEs which are nonlinear in the $\s_{ij} (u,v)$ and appear to be untractable.

\vfill\eject

\end{document}